# Herausforderungen und Chancen für die Lithiumgewinnung aus geothermalen Systemen in Deutschland

Valentin Goldberg[a,*], Tobias Kluge[b], Fabian Nitschke[a]

[*] Korrespondierender Autor: valentin.goldberg@kit.edu

[a] Institut für Angewandte Geowissenschaften, Professur Geothermie und Reservoir-Technologie, Karlsruher Institut für Technologie, Adenauerring 20b, 76131 Karlsruhe

[b] Institut für Angewandte Geowissenschaften, Professur Geochemie und Lagerstättenkunde, Karlsruher Institut für Technologie, Adenauerring 20b, 76131 Karlsruhe

## 1 Zusammenfassung

Die hier vorgestellte Arbeit bietet einen Ansatz, den technologischen Stand der Lithiumgewinnung aus geothermalen Wässern in Deutschland basierend auf aktuellen wissenschaftlichen Studien abzuschätzen, sowie die damit verbundenen Herausforderungen zu bewerten. Hierzu werden der prognostizierte weltweite Lithiummarkt, sowie der wegen des geplanten Aufbaus einer heimischen Batteriezellenfertigung enorm steigende deutsche Bedarf dargestellt. Aus der stark wachsenden Nachfrage, einer kompletten Abhängigkeit von schlecht diversifizierten Überseequellen und den dortigen Abbaumethoden mit negativen Umweltauswirkungen, lässt sich eine hohe strategische Bedeutung einer möglichen Binnenquelle ableiten.

Insbesondere werden in der Studie unterschiedliche Technologien zur Lithiumextraktion aus Thermalwässern wie Flüssig-Flüssig-Extraktion, anorganische Sorptionsmittel, elektrochemische Methoden und Membrantechnologien hinsichtlich ihrer Anwendbarkeit in geothermischen Systemen verglichen und bewertet. Basierend auf dem Technologievergleich und dem heutigen Ausbau der Geothermie in Deutschland wurden unterschiedliche Szenarien für die extrahierbare Menge an Lithiumkarbonat aus geothermalen Systemen in Deutschland und der französischen Seite des Oberrheingrabens berechnet. So lassen sich unter Berücksichtigung aller zurzeit aktiver Bohrungen eine gesamte Extraktionskapazität von ca. 4.000 Tonnen Lithiumkarbona pro Jahr prognostizieren.

Eine große Anzahl an Extraktionsmethoden wurde im Labor validiert. Eine Skalierung zu einem industriellen Prozess fehlt bisher. Der Technologie-Reifegrad beträgt für alle evaluierten Methoden maximal TRL 6. Dabei bleibt die tatsächliche Umsetzung der Extraktion die zentrale Herausforderung mit zahlreichen standortspezifischen Hürden. Das nutzbare Volumen des geothermalen Fluids und die gegenläufig zu optimierenden Parameter übertägige Verweilzeit und Extraktionseffizienz sind die Schlüsselparameter für das Prozessdesign. Sie kontrollieren die verfahrenstechnisch zu bewältigenden Volumenströme und damit die Größe der Reaktionsgefäße. Die für die Extraktion potentiell nötigen Anpassungen der Betriebsparameter können zu einem Anstieg des Scalingpotentials führen. Der verfahrenstechnische Einsatz von Säuren oder Basen erhöht zusätzlich die Korrosivität des Fluids. Eine für den Rohstoffabbau normalerweise essentielle Bewertung der Ressource bezüglich ihrer Größe und der Nachhaltigkeit ihrer Bewirtschaftung fehlt bisher. Um die insgesamt großen Potentiale dieser Technologie in Zukunft nutzen zu können, müssen diese zentralen Fragen geklärt werden. Dafür sollte die Kooperation von Industrie und Wissenschaft verstärkt werden.

## 2 Einleitung

Die Energiewende und die damit einhergehende Elektrifizierung erfordert die Verfügbarkeit einer Vielzahl an strategischen mineralischen Rohstoffen. Eines der zentralen Metalle, das von der EU als kritischer Rohstoff definiert wurde, ist Lithium (Europäische_Kommission 2020). Bis vor wenigen Jahren waren die Hauptanwendungen für das Alkalimetall der Einsatz in der Keramik- und Glasindustrie, Schmiermittel oder für Aluminiumlegierungen, wohingegen heute der größte Teil für die Herstellung von Lithium-Ionen-Akkus verwendet wird (Liu, Zhao, and Ghahreman 2019; Schmidt 2017). Aufgrund ihrer höheren Energiedichte, den niedrigen Entladungsraten, langer Lebenszeit und schnellem Ladeprozess im Vergleich zu Nickel-Metallhydrid oder Bleiakkumulatoren (Lee et al. 2011; Schmidt 2017; Kavanagh et al. 2018), fanden sie zunächst Anwendung im Bereich von portablen Geräten wie Kameras oder Mobiltelefonen. Heute ist der Einbau in Elektroautos der wichtigste Wachstumsmarkt. Um mit den Reichweiten der Verbrennungsmotoren konkurrieren zu können, werden große elektrische Speicherkapazitäten in der Größenordnung 60 – 100 kWh benötigt. Dies resultiert in entsprechenden Batteriegrößen von 600 - 800 kg und Volumina von 0,4 – 0,6 m$^3$ (Srinivasan et al. 2008; Lee et al. 2011; Tesla Inc. 2019). Der Anteil reinen Lithiums in einer Batterie für ein Elektroauto beträgt dabei ca. 12 kg in Abhängigkeit der Kapazität (Tesla Inc. 2019). Aufgrund der hohen Nachfrage und der daraus resultierenden Preisentwicklung werden global neue potentielle Lithiumquellen evaluiert, auch um umweltschonendere Alternativen zum konventionellen Bergbau oder der Extraktion aus den Wässern von Salzseen zu finden. Zurzeit wird die Lithiumgewinnung aus geothermalen Wässern als neue, nachhaltigere Abbaumethode, die auch einen Abbau in Deutschland möglich erscheinen lässt, diskutiert. Die vorliegende Arbeit betrachtet die Lithiumgewinnung aus geothermalen Wässern in Deutschland als mögliche Zukunftstechnologie und stellt technische Hürden für eine nachhaltige und ökologische Rohstoffgewinnung dar.

Eine erfolgreiche Umsetzung würde Deutschland unabhängiger von globalen Rohstoffmärkten machen und böte die Möglichkeit, globale Preisschwankungen zu puffern und Versorgungsschwierigkeiten abzufedern. Weiterhin könnte eine kombinierte stoffliche und energetische Nutzung eines geothermalen Reservoirs zu einer Symbiose mit positiven wirtschaftlichen Effekten für die Geothermiebranche führen und gleichzeitig die Rohstoffgewinnung nachhaltiger gestalten. Mit der Kombination aus erneuerbarer Energie- und Rohstoffproduktion ohne großen Flächenbedarf und negative Umweltbeeinflussung, stellt dies eine große Chance für einen modernen und zukunftsorientierten Umgang mit Ressourcen in Europa und Deutschland dar.

## 2.1 Markt und Prognosen

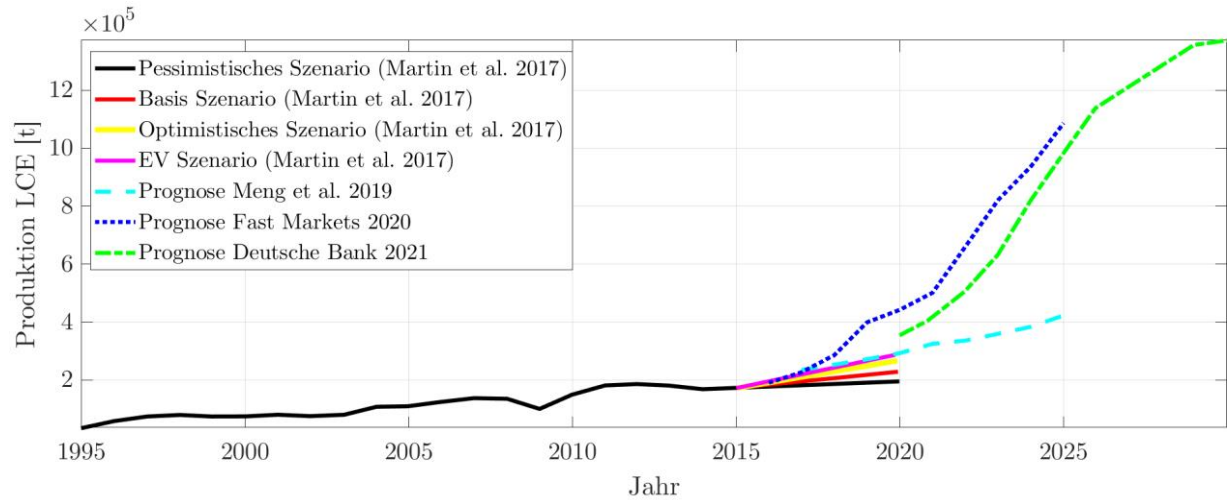

*Abbildung 1: Globale jährliche Lithiumproduktion als Lithiumkarbonatäquivalente (LCE) bis 2015 und Darstellung verschiedener Szenarien bis 2030 (Datenherkunft:* (Martin et al. 2017; Meng et al. 2021; Adams 2020; Jones et al. 2021)*).*

Die Entwicklung der globalen jährlichen Lithiumproduktion, in Abbildung 1 dargestellt als Lithiumkarbonat-Äquivalente (LCE), zeigt in den letzten 25 Jahren einen starken Aufwärtstrend von ca. 40.000 t im Jahr 1995 auf 420.000 t im Jahr 2020. Bis zum Ende 2021 wird bereits eine Produktion von 520.000 t erwartet. Ein deutlicher Einbruch der Produktion, verursacht durch die Weltwirtschaftskrise, ist für das Jahr 2008 festzustellen. Im Jahr 2010 hatte sich der Markt aber bereits wieder erholt und neue Höchstwerte erreicht. Die politische Wende hin zur Elektromobilität, wie sie in großen Märkten wie Europa und China der Fall ist, führt seit 2015 zu einem deutlich stärkeren Wachstumsanstieg als in den Jahrzehnten zuvor. In den letzten Jahren nahm der Ausbau der Elektromobilität so stark zu, dass die Prognosen des optimistischen „Electrical Vehicle Scenario" (EV Szenario) von 2017 und die Prognosen der Szenarien von 2019 allesamt 2021 deutlich übertroffen wurden (Martin et al. 2017; Meng et al. 2021). Hält dieses Wachstum an, so ist eine Verdreifachung der Lithiumproduktion bis 2025/2026 prognostiziert (Adams 2020; Jones et al. 2021; Al Barazi et al. 2021).

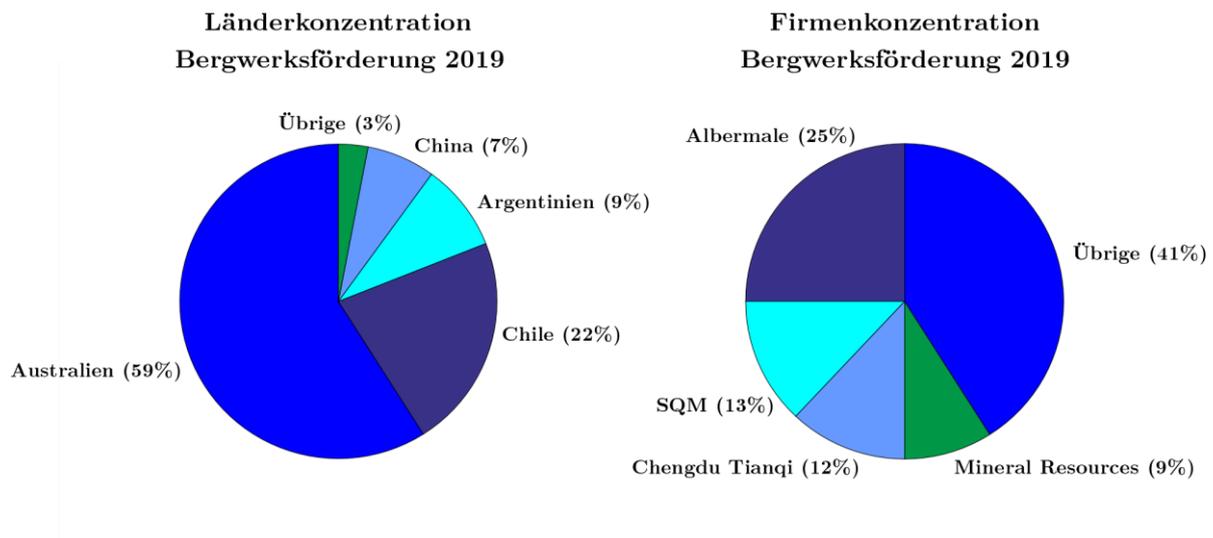

*Abbildung 2 Länder- und Firmenkonzentrationen der Lithium-Bergwerksförderung weltweit* (Al Barazi et al. 2021)

Die globale Lithiumförderung konzentriert sich im Wesentlichen auf drei Länder und vier Firmen. (Abbildung 2). Der größte Lithiumproduzent ist mit Abstand Australien, wo Lithium bergmännisch aus dem Mineral Spodumen gewonnen und hauptsächlich für den Export nach China produziert wird (Schmidt 2017; Al Barazi et al. 2021). Zweit- und drittgrößter Produzent sind Chile und Argentinien. Hier wird Lithium aus salzhaltigen, oberflächennahen Wässern über Evaporation und Fällung gewonnen. Für den europäischen Importbedarf wird eine derart starke Länderkonzentration allgemein kritisch betrachtet, da sich die Produktion auf wenige Länder beschränkt, darunter kein nennenswerter europäischer Standort. Weiterhin ist auch die auf dem Markt frei verfügbare Menge durch langfristige Lieferverträge stark reduziert (Al Barazi et al. 2021). Mit dem geplanten Ausbau der Batteriezellenfertigung an 9 Standorten in Deutschland, sollen jährlich mittelfristig 55 GWh Gesamtleistung produziert werden. Bei vollständigem Ausbau soll diese bis zu 215 GWh erreichen. Daraus würde ein jährlicher Lithiumbedarf von 7.000 t – 28.000 t (entspricht 37.000 – 149.000 t LCE) resultieren (Al Barazi et al. 2021). Um diesen Bedarf zu decken würde Deutschland jährlich 3 % - 15 % der von der Deutschen Bank global prognostizierten LCE Produktion in 2025 benötigen, ohne aktuell über eine heimische Produktion zu verfügen.

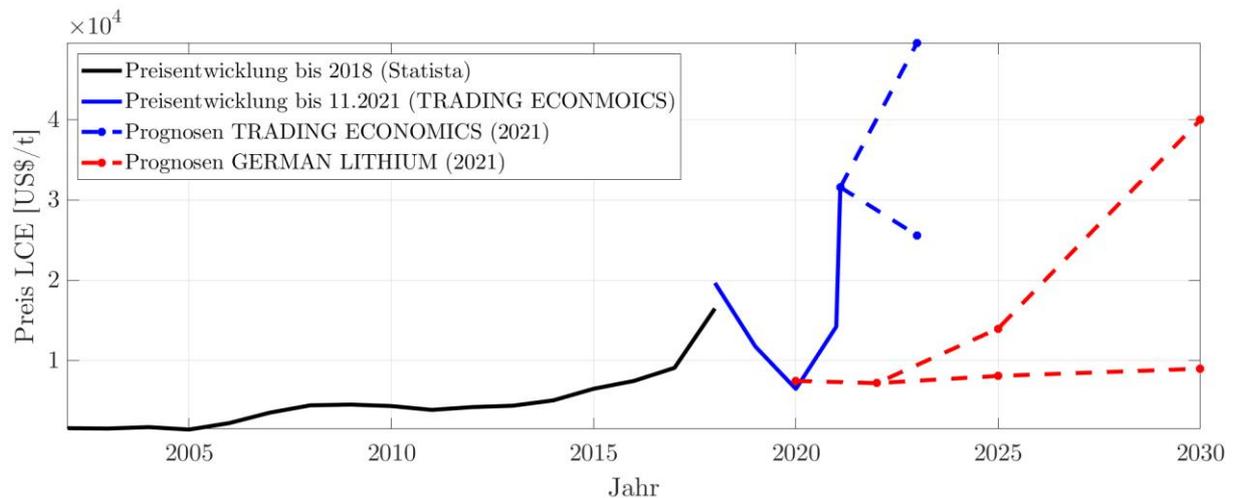

*Abbildung 3 Preisentwicklung für LCE von 2000 – 2021 und Prognosen bis 2030 (Datengrundlage:* (Hohmann 2021; Economics 2021; GERMANLITHIUM 2021)*)*

Die Veränderung von Angebot und Nachfrage hatte auch starke Auswirkungen auf die Entwicklung der Lithiumpreise (Siehe Abbildung 3). Die Produktionskosten sind dabei je nach Lagerstätten- und Abbau-Typ stark unterschiedlich. Auch wenn die Abbaukosten der lithiumhaltigen Erzlagerstätten im Vergleich zum Marktpreis von Lithium gering sind (250 – 400 US$/t), liegen die Kosten der Aufbereitung, bis eine Batteriequalität erreicht ist, höher als die Gewinnung von Lithiumkarbonat aus Sole (Schmidt 2017). Die Produktionskosten für eine Tonne Lithiumkarbonat aus dem Lithiumbergbau mit der höchsten Produktion weltweit (Greenbushes, Australia) beträgt ca. 4.500 US$ und liegt damit über den Kosten der Lithiumkarbonatherstellung aus den Salzseen in der Atacamawüste mit 2.500 – 3.000 US$ (Schmidt 2017).

Seit 2002 sind die Preise für Lithiumkarbonat von 1.590 US$/t auf über 30.000 US$/t im Jahr 2021 gestiegen (Abbildung 3). Der durch den starken Preisanstieg zwischen 2016 und 2018 bedingte Ausbau der Produktion vor allem in Australien führte zu einem Überangebot, das in Kombination mit einer verzögerten Nachfrage zu einem Preisverfall 2020 führte (Al Barazi et al. 2021). Von diesem Tief hat sich der Lithiumpreis erholt und im 4. Quartal 2021 einen neuen absoluten Höchstwert erreicht. Unterschiedliche Szenarien prognostizieren ein weiteres Wachstum des Lithiummarktes (Schmidt 2017; Martin et al. 2017; Meng et al. 2021; Adams 2020; Jones et al. 2021). In der Studie der Deutschen Rohstoffagentur DERA (Schmidt 2017) wird von einem jährlichen Zuwachs der Nachfrage zwischen 7,3 und 12,8 % ausgegangen. Im minimalen Fall von 7,3 % ergibt sich für das konservative Modell ein Angebotsüberschuss von ca. 22 %, wodurch ein Anstieg des Preises für Lithiumkarbonat unrealistisch wäre. Tritt der maximale Fall von 12,8 % Nachfragesteigerung ein, ist mit einem Angebotsdefizit von ca. 18 % zu rechnen, das dementsprechend eine Preissteigerung zur Folge hätte (Schmidt 2017). Vergleicht man die Prognosen des Lithiummarktes von 2017 mit dem tatsächlich eingetretenen Marktvolumen 2021, zeichnet sich ab, dass die konservativen Modelle die Situation deutlich unterschätzen. Für die Preisentwicklung gibt es unterschiedliche Voraussagen (Abbildung 3): 2020 wurde eine Preisentwicklung zwischen 8.000 und 14.000 US$/t LCE für 2025 aufgrund der temporär niedrigen Preise prognostiziert (GERMANLITHIUM 2021, basierend auf Daten von: Citigroup 2020, UBS/Forbes 2020, Penisa et al. 2020 und Seeking Alpha 2020). Studien von 2021 sagen dagegen für 2023 Preise zwischen 25.600 und 49.600 US$/t LCE (Economics 2021) und für 2030 zwischen 9.000 und 40.000 US$/t LCE (GERMANLITHIUM 2021) voraus und zeigen basierend auf den aktuellen Werten eine deutlich drastischere Preissteigerung auf.

## 2.2 Lithium in geothermalen Systemen

Lithium kommt neben den konventionellen Lagerstätten auch in Solen und Thermalwässern in sehr unterschiedlichen Konzentrationen vor. Welcher Lithiumgehalt und welche Fließrate als ausreichend für eine ökonomische Extraktion ist, hängt vom Stand der Technik, dem Extraktionsverfahren (s. Kapitel 3) und dem Weltmarktpreis ab. Um im Folgenden eine Kategorisierung zu ermöglichen, sprechen wir von erhöhten Konzentrationen, wenn das Fluid mehr als 1 mg/L Lithium enthält, von hohen Konzentrationen jenseits von 100 mg/L. Meerwasser enthält zum Vergleich nur 0,18 mg/L Lithium (Stoffyn-Egli and MacKenzie 1984). Erhöhte Lithiumkonzentrationen >1 mg/L finden sich in Europa in einigen ausgewählten Regionen: neben dem Oberrheingraben und dem Norddeutschen Becken auch in Cornwall, Nordengland, dem französischen Zentralmassiv, dem Pariser Becken und der westlichen Apenninenregion (Stringfellow and Dobson 2021). Über die Konzentration von Lithium in geothermalen Fluiden anderer europäischer Regionen wurde bislang wenig publiziert. Daher kann das Potential hier nicht abschließend bewertet werden. Ausgesprochen hohe Werte >100 mg/L wurden in Tiefenwässern des Oberrheingrabens und dem Norddeutschen Becken gemessen (Regenspurg, Milsch, and Schaper 2015; Sanjuan, Millot, et al. 2016). An den aktiven Geothermieanlagen Insheim, Landau und Bruchsal auf deutscher Seite und Soultz-sous-Forêts sowie Rittershoffen auf französischer Seite im Oberrheingraben wurden Konzentrationen von 160-190 mg/L Lithium gemessen. Das Fluid aus der Forschungsbohrung Groß-Schönebeck im Norddeutschen Becken enthält etwa 215 mg/L. Diese Werte reichen an die Untergrenze des Konzentrationsbereichs, der Salare heran (ca. 300-1600 mg/L; Stringfellow und Dobson, 2021) und können bei hohen Fließraten eine wirtschaftliche Relevanz erreichen. Hohe Lithiumkonzentrationen >100 mg/L wurden in Deutschland auch in anderen geologischen Formationen nachgewiesen: z.B. in Kohlenwasserstoffbohrungen im oberen Muschelkalk des südwestlichen Molassebeckens (Stober 2014). Des Weiteren, wurden auch in den flachen Bohrungen an der Taunusrandstörung erhöhte Konzentrationen gemessen, z.B. in Bad Nauheim mit ca. 10 mg/L (Kirnbauer 2008; Loges et al. 2012).

Das Konzept der Lithiumförderung aus tiefen geothermalen Vorkommen sieht vor, die Ressource analog zur geothermischen Produktion über Tiefbohrungen zu erschließen. Aktuelle Bestrebungen sehen eine Co-Nutzung bestehender Thermalwasserproduktion vor. Die Standorte für Tiefe Geothermie in Deutschland verteilen sich im Wesentlichen auf drei geologische Großstrukturen, den Oberrheingraben (ORG), das Bayrische Molassebecken (BMB) sowie das Norddeutsche Becken (NDB). Insgesamt wird in Deutschland Stand 2021 an 42 Standorten Thermalwasser zur energetischen Nutzung aus Tiefen >400 m produziert (Abbildung 4, (Agemar et al. 2014; Agemar, Weber, and Schulz 2014)). 30 Anlagen dienen ausschließlich der Wärmegewinnung, an drei Anlagen wird Strom produziert und an neun Standorten wird Wärme und Strom kombiniert genutzt. Insgesamt sind etwa 350 MW Wärmeleistung und 47 MW elektrische Leistung installiert. Somit können in Deutschland bei Verfügbarkeit aller Anlagen kumuliert etwa 3150 L/s Tiefenwasser produziert werden (Dilger et al. 2021). Beschränkt man die Standorte auf die produzierenden Bohrlöcher des ORG und des NDB, wo erhöhte Li-Konzentrationen bekannt sind, dann verringert sich das kumulierte produzierte Volumen auf nur noch etwa 270 L/s. Unter Berücksichtigung der französischen Anlagen in Soultz-sous-Forêts und Rittershoffen werden insgesamt 370 L/s gefördert.

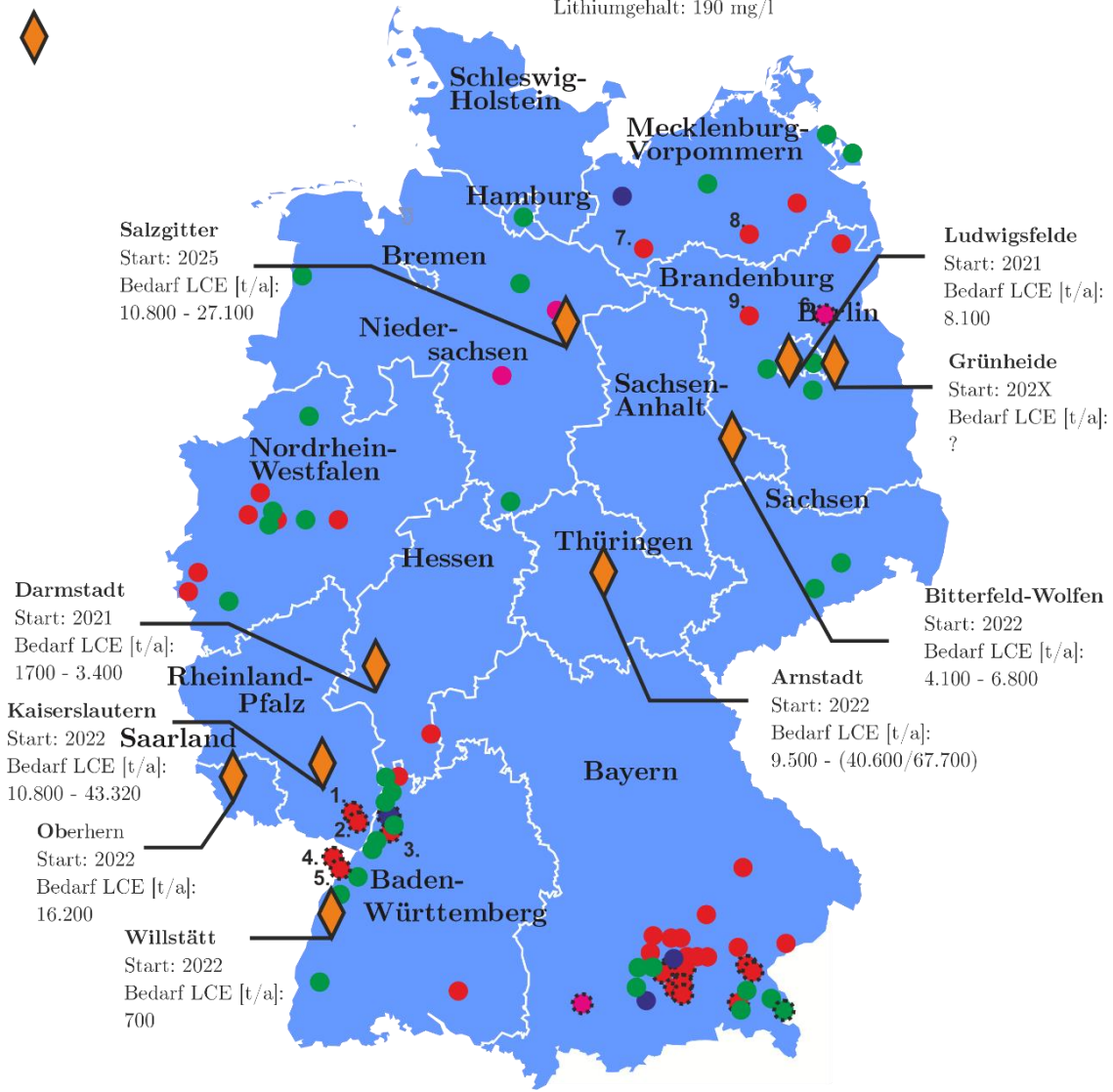

*Abbildung 4 Geothermiestandorte und geplante Batteriezellfertigungen in Deutschland. Die Karte zeigt die Lithiumgehalte und Fließrate der für die Rohstoffextraktion interessanten Standorte sowie den prognostizierten Bedarf der Zellfertigungen. Der Bedarf wurde berechnet basierend auf den geplanten jährlichen Kapazitäten. Datenquellen:* (Al Barazi et al. 2021; Agemar et al. 2014; Agemar, Weber, and Schulz 2014).

### 2.2.1 Thesen zur Lithiumherkunft

Für die Exploration, sowie zur Bewertung der Ressourcen und der Entwicklung eines nachhaltigen Extraktionskonzepts, ist eine Kenntnis der Lithiumherkunft und ein Modell zur Fluidgenese elementar. Beide Aspekte sind für Fluide in Deutschland Teil aktiver Forschung und erlauben daher noch keine endgültigen Aussagen.

Erhöhte Lithiumkonzentrationen ≥1 mg/L lassen sich nicht grundsätzlich mit tieferen Bohrungen erzwingen, da die Konzentration nicht per se mit der Tiefe der Bohrung korrelieren. Zum Beispiel liegen die Lithiumkonzentrationen der Thermalwässer im Oberen Jura-Aquifer des bayrischen Molassebeckens nur bei <0,1 mg/L, trotz einer Reservoirtiefe von z.T. über 2000 m (Stober, Wolfgramm, and Birner 2014). Diese Wässer zeigen weiterhin geringe Chloridgehalte, die nur im westlichen Bereich des Molassebeckens 2 g/L überschreiten (Stober, Wolfgramm, and Birner 2014). Im Gegensatz dazu werden in der oberflächennahen Bohrung von Bad Nauheim von nur 180 m bereits Konzentrationen von ca. 10 mg/L erreicht (Kirnbauer 2008; Loges et al. 2012). Experimente zur Gesteins-Wasser-Wechselwirkungen zeigen, dass bestimmte Reservoirgesteine des Oberrheingrabens oder des permischen Rotliegends, erhöhte Potentiale für eine Abgabe von Lithium an das Formationswasser haben (Drüppel et al. 2020; Regenspurg, Milsch, and Schaper 2015). Des Weiteren zeigen die experimentellen Daten bzgl. Fluidwechselwirkungen mit Granit einen signifikanten Zusammenhang zwischen dem Lösungspotential des Lithium und der Salinität des Fluids (Drüppel et al. 2020).

Zur Genese der lithiumhaltigen Fluide im Oberrheingraben und dem norddeutschen Becken gibt es unterschiedliche Hypothesen:

   a) Initiale Lithiumanreicherung während der Evaporation von Meerwasser. Dieses Fluid verbleibt nach der initialen Phase in der jeweiligen Schicht und wird ggf. durch andere Wässer verdünnt. Eine Lithiumanreicherung wurde bei der Evaporation von Meerwasser zumindest in Experimenten beobachtet (z.B.(Bąbel and Schreiber 2014)).
   b) Verdunstung von Meerwasser mit Zunahme der Salinität. Wechselwirkung des Fluids mit dem entsprechenden Reservoirgestein unter erhöhter Temperatur nach Migration in entsprechende Aquiferschichten. Dabei kommt es zur Alteration der Feldspäte und Freisetzung von Lithium ins Fluid (Drüppel et al. 2020). Dieses an Lithium angereicherte Fluid verbleibt dann in den entsprechenden Schichten und wird ggf. durch Mischung verdünnt.
   c) In einem komplexeren Ansatz entsteht ein thermales Mischwasser, dass Anteile aus verschiedenen Tiefen und Reservoiren integriert (Burisch et al. 2018). Endglieder sind Meerwasser, salinare Wässer aus Halitauflösung und die meteorische Komponente. Die salinaren Wässer wechselwirken über einen langen Zeitraum (viele 100 ka) und bei erhöhten Temperaturen mit dem Reservoirgestein und reichern dabei u.a. Lithium an. Bislang ist nicht bekannt, ob abgesehen von Granit, Rotliegend- und Zechsteinlagen auch andere Gesteine und sedimentäre Schichten signifikante Lithiumkonzentrationen aufweisen können.

## 3 Extraktionstechnologien

Für die Rohstoffgewinnung aus Thermalwässern ist eine Lithiumextraktion anzustreben, die so wenig wie möglich in das chemische Gleichgewicht des Thermalwasserstroms im Kraftwerk und Reservoir eingreift. Hierbei kommt eine Vielzahl an unterschiedlichen Ansätzen sowie Kombinationen in Frage. Für die Betrachtung unterschiedlicher Extraktionstechnologien ist es wichtig, deren Selektivität, das Verhältnis von Materialaufwand zu Lithiumausbeute, die chemischen Eigenschaften des benötigten Materials und den Energieeinsatz zu bewerten.

Bereits frühere Studien zeigten den Mangel an konsistenten Daten zu den jeweiligen Methoden auf, was eine Herausforderung für einen Technologievergleich darstellt (Battistel et al. 2020; Stringfellow and Dobson 2021; Liu, Zhao, and Ghahreman 2019). Ziel des Vergleichs ist es daher herauszuarbeiten, welche Technikaspekte besondere Herausforderungen für die jeweilige Technologie beim Einsatz in Geothermiekraftwerken darstellen. Eine Zusammenfassung der Technologien ist in Tabelle 1 dargestellt. Für die Abschätzung des Technologie-Reifegrads wurden ausschließlich wissenschaftlich begutachtete Studien berücksichtigt. Firmen-Kommunikation oder Trivialliteratur wurden nicht beachtet.

*Tabelle 1 Verschiedene direkte Extraktionsverfahren für Lithium aus Thermalwässern. Die Tabelle gibt einen Überblick über Vor- und Nachteile der Technologien, Betriebsparameter während des Extraktionsprozesse, sowie eine Abschätzung des Technologie-Reifegrads (TRL).*

| Methode | Technologien | Vorteile | Nachteile | Exkrationsparameter | TRL |
|---|---|---|---|---|---|
| Flüssig-Flüssig Extraktion | Tributylphosphate<br>Kronenether<br>Ionische Flüssigkeiten | Kurze Beladungszeit<br><br>Extraktionsmittel recyclebar | Chemikalenverbrauch für Rücklösung und Regeneration<br><br>Großer Platzbedarf für die benötigten Fluidströme<br><br>Hohe Kosten (Bei Kronenethern)<br><br>Korrosion durch Säuren und Baseneinsatz | Extraktionseffizienz: 40 - 90 %<br>Reaktionszeiten: 10 - 50 Min.<br>Verhältnis Lösungsmittel - Sole: 1:1 - 1:2,5 | 3 - 4 |
| Anorganische Sorbentia | Mangan- & Titanadsorber (Ionensiebe) | Hohe Beladungskapazitäten | Adsorber werden durch Einsatz und Regeneration abgebaut<br><br>pH-Wert Änderung kann zu Scaling führen<br><br>Lange Beladungszeit<br><br>Korrosion durch Säuren und Baseneinsatz | Extraktionseffizienz: 40 - 95 %<br>Beladungskoeffizient: 20 - 60 mg/g<br>Idealer pH-Wert bei Extraktion: 10 - 13<br>Reaktionszeiten: 60 Min. -120 h | 4 - 5 |
| | Aluminiumhydroxid | Geringer Chemikalienverbrauch<br>Gute Recyclebarkeit | Niedrige Beladungskapazität<br>Hoher Frischwasserbedarf | Extraktionseffizienz: 50 - 90 %<br>Beladungskoeffizient: < 8 mg/g<br>Idealer pH-Wert bei Extraktion: 7<br>Reaktionszeiten: 60 Min. - 10 h.<br>Wasserbedarf für Stripping: ~100 fache Menge des Adsorbers | (6) |
| Elektrochemische Methoden | Ionenpumpe<br>Elektrodialyse | Kurze Reaktionszeit<br>Kein Chemikalienverbrauch<br>Prozesse verwandt zu den in einer Batterie<br>Arbeiten auch bei niedrigen Lithiumgehalten effizient | Zusätzlicher Energiebedarf<br>pH-Effekte des Prozesses können zu Scaling führen<br>Können nur bis zu einem Lithiumgehalt von 350 mg/L eingesetzt werden | Beladungskoeffizient der Arbeitselektrode: 30 - 40 mg/g<br>pH-Wert Schwankungen durch Methode: +/- 2<br>Reaktionszeiten: < 20 Min.<br>Zusätzlicher Strombedarf: 1 – 60 Wh/mol Lithium | 3-4 |

| Membrantechnologien | Nanofiltration / Membrandestillation | Energieeffizient  Gut integrierbar in geothermische Kreisläufe  Aus der Wasseraufbereitung bekannt und industriell verwendet  Zusätzliche Möglichkeit der Wassergewinnung | Integrität der Membranen ist durch Scaling gefährdet  Ohne zusätzliche Verfahren wird Lithium nur relativ angereichert und nicht selektiv extrahiert | | 4 |
|---|---|---|---|---|---|
| | Supported Liquid Membran | | Sehr hohe pH-Werte für optimale Bedingungen benötigt  Großer Chemikalieneinsatz und hohe Wahrscheinlichkeit für Scaling | Extraktionsrate: 90 %  Reaktionszeit: 120 Min.  pH-Wert: 9,5 - 12,5 | 3 - 4 |
| | Kombination mit Ionensieben | Hohe Beladungskapazitäten  Schnellere Kinetik als Adsorber alleine | Adsorber werden durch Regeneration abgebaut  pH-Wert Änderung kann zu Scaling führen  Lange Beladungszeit  Korrosion durch Säuren und Baseneinsatz | Beladungskoeffizient: 30 mg/g  Reaktionszeit: < 60 Min.  →Rest; Analog zu Mangan-& Titanadsorbern | |

## 3.1 Flüssig-Flüssig Extraktion

Bei der Flüssig-Flüssig Extraktion werden gelöste Stoffe eines flüssigen Ausgangsmediums in ein ebenfalls flüssiges Lösungsmittel (engl. *Solvent*, daher auch die Bezeichnung Solventextraktion) überführt (Liu, Zhao, and Ghahreman 2019). Grundvoraussetzung dafür ist, dass sich das Ausgangsmedium nicht mit dem Lösungsmittel mischt, jedoch bei Kontakt den zu extrahierenden Stoff austauscht. Organische Lösungsmittel können relevante Mengen Lithiumchlorid aufnehmen und auch eine Selektivität gegenüber anderen Kationen zeigen (Liu, Zhao, and Ghahreman 2019; Stringfellow and Dobson 2021). Vielversprechende Ansätze für die Gewinnung von Lithium aus hochsalinaren Wässern sind unter anderem die Verwendung von Tributylphosphaten (TBP), verdünnt mit Methylisobutylketon oder Kerosin (Liu, Zhao, and Ghahreman 2019; Xiang et al. 2017; 2016; Nguyen and Lee 2018; D. Shi et al. 2018; Yu et al. 2019), die Anwendung von Kronenethern (Liu, Zhao, and Ghahreman 2019; Swain 2016; Z. Zhang et al. 2021) oder der Einsatz von ionischen Flüssigkeiten (Liu, Zhao, and Ghahreman 2019; Park et al. 2014; C. Shi et al. 2017).

Bei der Extraktion mittels TBP wird zusätzlich zu den Lösungsmitteln eine Koextraktionsreagenz verwendet wie z.B. $FeCl_3$. Aus dem $FeCl_3$ bildet sich in chloridreichen Wässern zusammen mit den Lösungsmitteln ein Komplex, der über einen Kationenaustauschprozess das Lithium bindet. Für diesen Prozess müssen die beiden Flüssigkeiten samt der Koextraktionsreagenz ein chemisches Gleichgewicht erreichen. Im darauffolgenden Schritt müssen die beiden Phasen wieder getrennt werden, bevor im Anschluss mittels hochkonzentrierter Säure die Rücklösung des Lithiums aus dem Lösungsmittelkomplex erfolgen kann. Abschließend werden Lösungsmittel und Koextraktionsreagenz unter der Verwendung von Laugen regeneriert (Xiang et al. 2016; Nguyen and Lee 2018; D. Shi et al. 2018; Liu, Zhao, and Ghahreman 2019).

Die Funktion von Kronenethern basiert auf dem „Hard and Soft Acids and Bases"-Konzept (HSAB / Pearson-Konzept). Demnach binden starke Säuren bevorzugt an starke Basen und schwache Säuren an schwache Basen. Bei Kronenethern fungiert Sauerstoff als Donoratom, das eine starke Base bildet und eine sehr gute Reaktivität mit gelösten Alkalimetallkationen wie Lithium zeigt, die eine starke Säure darstellen. Die Kationen werden dadurch im Zentrum der Kroenetherstruktur gebunden (Liu, Zhao, and Ghahreman 2019; Swain 2016). Nach Aufnahme des Lithiums wird die Etherphase von der Wasserphase getrennt und das Lithium kann mit Säure aus den Ethern rückgelöst werden (Z. Zhang et al. 2021).

Als ionische Flüssigkeiten werden Salze bezeichnet die aus organischen Kationen und organischen oder anorganischen Anionen bestehen und Schmelztemperatureb kleiner als 100 °C aufweisen (C. Shi et al. 2017; Park et al. 2014). Für die gezielte Rohstoffextraktion können sie direkt als Extraktionsmittel dienen oder in Kombination mit anderen organischen Lösungsmitteln als Co-Extrator (Liu, Zhao, and Ghahreman 2019; Stringfellow and Dobson 2021; C. Shi et al. 2017; Swain 2017). Auch bei diesem Ansatz muss die Phase der ionischen Flüssigkeiten zunächst vom ursprünglich lithiumführenden Fluid separiert und im Anschluss für die Auslösung des Lithiums mit Säure behandelt werden. Das Lösungsmittel selbst wird mit Natronlauge regeneriert (C. Shi et al. 2017).

Bei der Flüssig-Flüssig Extraktionsanwendung wurden im Labor aus wässrigen Lösungen bis zu 90 % des enthaltenen Lithiums extrahiert (Liu, Zhao, and Ghahreman 2019; Yu et al. 2019).

Durch Extraktion mittels TBP wurden 90 %, Extraktionseffizienz jedoch erst nach fünf in Reihe geschalteten Extraktionsdurchgängen erreicht. Bei einmaligem Kontakt lag die Rate bei ca. 40 % (Yu et al. 2019). Zum größten Teil lassen sich die diskutierten Lösungsmittel nach der Lithiumgewinnung wieder regenerieren und weiterverwenden. Hierbei werden jedoch große Mengen Säure für die Mobilisierung des Lithiums benötigt sowie in den meisten Ansätzen auch Natronlauge zur Regeneration. Der Verbrauch der Materialien ist entsprechend in eine Wirtschaftlichkeitsbetrachtung mit einzubeziehen. Auch der Verbleib der Säuren und Laugen nach der Behandlung muss mit Blick auf den ökologischen Fußabdruck betrachtet werden. Zudem kommt es sowohl beim Kontakt mit dem Trägerfluid, als auch bei der Aufbereitung im Anschluss zu geringen Verlusten der Lösungsmittel (Liu, Zhao, and Ghahreman 2019; C. Shi et al. 2017). Darüber hinaus stellen Säuren und Basen aufgrund ihrer Korrosivität hohe Ansprüche an das Material der Reaktionsgefäße und des Kraftwerks. Eine weitere Herausforderung sind die Volumenströme bei der Flüssig-Flüssig-Extraktion. Je nach Lösungsmittel variieren die Verhältnisse von geothermalem Fluid zu Lösungsmittel zwischen 1:1 und 1:2,5 (Yu et al. 2019; Z. Zhang et al. 2021; C. Shi et al. 2017; Garrett 2004; Flexer, Baspineiro, and Galli 2018). Dies erfordert für die hohen Fließraten in Geothermiekraftwerken einen entsprechenden hohen Lösungsmitteleinsatz. Die Lösungsmittel benötigen zudem eine ausreichende Kontaktzeit zum Ausgangsfluid von mindestens 10 bis 30 Minuten (Yu et al. 2019; C. Shi et al. 2017). In einem kontinuierlichen Durchflusssystem in einem Geothermiekraftwerk mit Fließraten von 80 L/s würde man ein Reaktionsvolumen von 48 – 144 m$^3$ benötigen, um entsprechende Verweilzeiten für das geothermale Fluid allein zu erreichen. Für die Extraktion muss zusätzlich das Volumen für das Lösungsmittel bereitgestellt werden. Ausgehend von einem 1:1 Verhältnis von Lösungsmittel und geothermalen Fluid benötigt man zwischen 96 m$^3$ und 288 m$^3$ für die Verweilzeiten von 10 bis 30 Min. Weiterhin bedarf auch das Trennen der beiden Phasen noch einmal einer Reaktionszeit in der gleichen Größenordnung. Durch den Einsatz von Zentrifugen kann diese verkürzt werden (Swain 2017; D. Shi et al. 2018; Yu et al. 2019).

Allgemein bieten die Flüssig-Flüssig Technologien ein hohes Potential und haben im Labor sehr gute Extraktionseffizienzen erzielt. Dies gilt auch für Extraktionsversuche aus natürlichen Wässern, was einem Technologie-Reifegrad (Technical Readiness Level) von 3-4 entspricht. Für den technischen Einsatz im Anlagenmaßstab im Rahmen der Behandlung von Thermalwässern eines Geothermiekraftwerks sollten jedoch unterschiedliche Parameter wie die Reaktionszeit, der benötigte Volumenstrom und der Chemikalieneinsatz optimiert werden, um einen wirtschaftlichen Einsatz zu ermöglichen. Auch für die in einem geothermischen Kreislauf benötigte Druck- und Temperaturhaltung sind Lösungen zu entwickeln.

## 3.2 Anorganische Sorbentia

Für die Verwendung anorganische Sorbentia zur selektiven Ionenabscheidung werden unterschiedliche Ansätze und Materialien erforscht, wie z.B. Titanoxide, Manganoxide, Aluminiumhydroxide oder Zeolithe. Dabei liegt der aktuelle Fokus der Forschungsprojekte darin, die Leistung der Adsorber mit Blick auf ihre Adsorptionskapazität und Wiederverwendbarkeit zu optimieren (Liu, Zhao, and Ghahreman 2019; Stringfellow and Dobson 2021). Das Grundprinzip beruht auf Ionenaustausch und Sorption. Bei den Prozessen wird der gelöste Rohstoff an einen Feststoff, der dem Fluid zugegeben oder von diesem durchströmt wird, gebunden. Nach dem Kontakt zwischen Sorptionsmittel und dem zu gewinnenden Rohstoff, muss dieser aus der beladenen Sorbens rückgelöst werden. Je nach Anwendung muss der Feststoff dafür zunächst abfiltriert werden oder die Durchflusszelle in der er sich befindet, direkt behandelt werden.

Manganoxide mit Spinellstruktur zeigen eine sehr gute Wirksamkeit (Liu, Zhao, and Ghahreman 2019; Weng et al. 2020; Stringfellow and Dobson 2021; Slunitschek, Kolb, and Eiche 2021). Diese werden unter Zugabe von Lithium synthetisiert, wodurch dieses in die Kristallstruktur eingebaut wird. Nach dem Herauslösen des Lithiums verbleibt eine Lücke weshalb man auch von (Lithium-)Ionensieben spricht (Weng et al. 2020; Liu, Zhao, and Ghahreman 2019). Die Bindung des Lithiums erfolgt über Redox-Reaktionen und Ionenaustausch an und im Kristallgitter (Weng et al. 2020; Liu, Zhao, and Ghahreman 2019). Dabei werden die Ionenaustausch-Reaktionen durch hohe pH-Werte begünstigt, was sich positiv auf die Adsorptionskapazität und benötigte Reaktionszeit auswirkt (Weng et al. 2020; Liu, Zhao, and Ghahreman 2019). Nach dem Kontakt des Sorptionsmittels mit dem lithiumhaltigen Fluid kann das Lithium von der Sorbens über eine Säurebehandlung gelöst werden (Weng et al. 2020; Ryu et al. 2019; Liu, Zhao, and Ghahreman 2019; Stringfellow and Dobson 2021). Die erreichten Adsorptionskapazitäten bei Tests in Meerwasser oder Sole liegen bei Manganadsorbern zwischen 20 mg/g und 60 mg/g (Weng et al. 2020; Liu, Zhao, and Ghahreman 2019; Ryu et al. 2016; Zandevakili, Ranjbar, and Ehteshamzadeh 2014). Die Kapazitäten > 30 mg/g wurden bei pH-Werten zwischen 10 und 13 erzielt (Liu, Zhao, and Ghahreman 2019). Hohe initiale Lithiumgehalte begünstigen die Extraktionseffizienz (Ryu et al. 2016). Mit den entsprechenden Mengen an Adsorbern konnten so bis zu 90 % des Lithiums aus unterschiedlichen Solen im Labor gewonnen werden (Zandevakili, Ranjbar, and Ehteshamzadeh 2014).

Die auf Titanoxiden basierenden Adsorber sind in ihrem Wirkungsmechanismus sehr ähnlich zu den Manganadsorbern. Sie zeigen jedoch eine höhere Stabilität über mehrere Adsorptionszyklen, sind dafür aber weniger selektiv (Stringfellow and Dobson 2021; Liu, Zhao, and Ghahreman 2019). Ihre Adsorptionsraten liegen im Bereich zwischen 20 und 30 mg/g, optimal im pH-Bereich 8 - 13 (Liu, Zhao, and Ghahreman 2019).

Die Lithiumgewinnung mit Aluminiumhydroxiden basiert auf dem Einbau von Lithium in den Oktaederlücken von geschichteten Aluminiumhydroxid-Mineralen wie Gibbsit unter der Bildung von Lithium-Aluminium-Doppelhydroxid-Chlorid (Stringfellow and Dobson 2021; Paranthaman et al. 2017; Wu et al. 2019; Isupov et al. 1999). Gleichzeitig, wird zwischen den einzelnen Schichten Chlorid gebunden, was eine Produktion von Lithiumchlorid bei der Rücklösung vereinfacht (Wu et al. 2019; Jiang et al. 2020). Im Vergleich zu den Mangan- und Titanadsorbern ist die Beladung deutlich geringer und erreicht weniger als 8 mg/g (Jiang et al. 2020; Isupov et al. 1999; Stringfellow and Dobson 2021). Für die Desorption benötigt man keine Säure, was den Verlust an Adsorber minimiert, dafür werden aber große Mengen Frischwasser im Verhältnis Wasser:Adsorber von 100:1 benötigt (Isupov et al. 1999). In einer Sensitivitätsanalyse (Jiang et al. 2020) wurde ein optimaler pH-Wert von 7 bestimmt, sowie eine Beladungsdauer von 60 Minuten für 50 % Beladung und 600 Minuten für das Erreichen des Gleichgewichts. Unter Berücksichtigung dieser Parameter konnten bis zu 90 % des Lithiums aus Solen im Labor gewonnen werden (Wu et al. 2019; Isupov et al. 1999).

Eine weitere Methode ist die Gewinnung von Lithium mittels natürlichen und künstlich synthetisierten Zeolithen. Diese erreichen über den Einbau mittels Ionenaustausch ebenfalls nur geringe Beladungskapazitäten von 5 mg/L und befinden sich noch im Anfangsstadium der Entwicklung für den Einsatz in Thermalwässern (Wísniewska et al. 2018; Stringfellow and Dobson 2021).

Die direkte Lithiumgewinnung mit Adsorbern zählt zu den am meisten erforschten Methoden. Darunter erzielen Mangan- und Titanadsorber die besten Beladungskapazitäten (Stringfellow and Dobson 2021; Liu, Zhao, and Ghahreman 2019). Dennoch ist der Materialeinsatz bei Beladungen von max. 40 mg/g Adsorber nicht zu vernachlässigen, wenn man Fluide mit Lithiumkonzentrationen von bis zu 200mg/L bei Fließraten von 80 L/s behandeln will. Um die teilweise langen Reaktionszeiten von 24 h - >120 h (Liu, Zhao, and Ghahreman 2019; Ryu et al. 2019; Zandevakili, Ranjbar, and Ehteshamzadeh 2014) für eine volle Beladung zu erreichen, benötigt man bei einem kontinuierlichen Volumenstrom große Reaktionsgefäße. Diese müssen in die Infrastruktur eines bestehenden Geothermiekraftwerks integrierbar sein. Eine weitere spezifische Herausforderung für diese Art der Adsorber ist, dass sie während der Desorption abgebaut werden. Dies erhöht den Materialverbrauch zusätzlich. Auch die Folgen eines Eintrags von Mangan oder Titan in den geothermischen Aquifer sind zu betrachten. Die Auflösungsraten schwanken dabei zwischen 1 - 7 % beim ersten Einsatz eines frisch synthetisierten Adsorbers und liegen danach bei ca. 1 % pro Zyklus (Zandevakili, Ranjbar, and Ehteshamzadeh 2014; Ryu et al. 2016). Für den optimalen Adsorptionsbereich erfordern die hohen pH-Werte Laugen zur Pufferung des Systems, die den Materialeinsatz zusätzlich erhöhen. Im basischen pH-Bereichen besteht außerdem die Gefahr von karbonatischen und silikatischen Scalings, die die Effektivität der Adsorber vermindern könnten. Der Materialverbrauch vergrößert sich zudem durch die benötigte Säure für die Rücklösung des Lithiums. Aluminiumhydroxid-Adsorber benötigen zur Pufferung des Systems Laugen bzw. Säuren, wenn der optimale pH Bereich von 7 erreicht werden soll. Für die Rücklösung genügt Wasser. Auch der Abbau der Adsorber ist gering. Jedoch stellt die geringe Beladung von maximal 8 mg/g und der hohe Wasserbedarf Nachteile dar. Um eine Extraktionsrate von 50 % erreichen, werden nach Jiang et al. (2020) eine ca. 60-minütige Reaktionszeit benötigt. Das Reaktionsvolumen muss mindestens den Durchfluss von 60 Minuten fassen können, was bei 80 L/s 288 m$^3$ entspräche. Bei Wässern mit einer Konzentration von 200 mg/L Lithium würden darin 57.600 g Lithium zirkulieren. Um davon 50 % innerhalb von einer Stunde zu extrahieren, benötigt man bei einer Gesamtkapazität von 8 mg/g ca. 7,2 Tonnen an Aluminiumhydroxid-Adsorbern sowie die 100-fache Menge, also 720 t, an Wasser für die Rücklösung. Daraus resultiert auch ein niedriger Lithiumgehalt in der extrahierten Lösung, der vor einer Weiterverarbeitung einer Aufkonzentration und Wasserrückgewinnung bedarf.

Zusammenfassend kann man sagen, dass die unterschiedlichen Adsorber und deren Funktion im Labor sehr gut erforscht sind und gute Ergebnisse erzielen. Versuche mit realen geothermalen Wässern zeigen das hohe Potential der Adsorber. Der Technologie-Reifegrad wird basierend auf der verfügbaren Literatur zwischen 4 und 5 beurteilt für Mangan- und Titanadsorber. Nach Berichten über erfolgreiche Feldtests (Rettenmaier et al. 2021; Stringfellow and Dobson 2021), wird der Reifegrad von Aluminiumhydroxiden höher eingeschätzt. Für den Einsatz mit realen Wässern in einem Geothermiekraftwerk muss für den Einsatz von Titan- oder Manganadsorbern die Stabilität verbessert werden, für

Aluminumhydroxid die Beladungskapazität und allgemein für die Adsorber die Kinetik, mit der die Lithiumaufnahme abläuft. Da die Verweilzeiten die Größe der benötigten Reaktionsgefäße maßgeblich beeinflussen, ist in Anbetracht der hohen Fließraten in den Geothermiekraftwerken eine schnelle Teilbeladung der Adsorber einer hochgradigen aber deutlich langsameren Beladung vorzuziehen. Weiterhin stellen sich bei langen Verweilzeiten neue Probleme der Temperaturhaltung sowie Oxidationen ein.

## 3.3 Elektrochemische Methoden

Das Grundprinzip der elektrochemischen Abscheidung von Lithium beruht vereinfacht auf der selektiven Anziehung von positiv geladenen Lithiumkationen an eine Arbeitselektrode unter Anlegen einer Spannung. Anionen und störende Elemente werden an eine Gegenelektrode gebunden werden (Liu, Zhao, and Ghahreman 2019; Battistel et al. 2020; Calvo 2019). Ähnlich zu den Adsorbern, können Mangan- und Titanoxide als Arbeitselektroden verwendet werden (Battistel et al. 2020; Stringfellow and Dobson 2021; Calvo 2019; Liu, Zhao, and Ghahreman 2019). Der große Vorteil der elektrochemischen Ansätze ist, dass die Bindung des Lithiums schnell abläuft und eine Rücklösung ohne Chemikalien möglich ist (Battistel et al. 2020). Die elektrochemischen Methoden zeigen schon bei niedrigen initialen Lithiumgehalten ab 7 mg/L eine gute Selektivität, sind jedoch auf einen Maximalwert von ca. 350 mg/L Lithium beschränkt (Battistel et al. 2020; Palagonia, Brogioli, and Mantia 2017). Die am häufigsten diskutierten Methoden sind die Ionenpumpe und die Elektrodialyse (Battistel et al. 2020; Stringfellow and Dobson 2021; Calvo 2019).

Beide Methoden verwenden eine Arbeitselektrode, die Lithium binden kann (Calvo 2019; Battistel et al. 2020). Deren Beladungskapazität gibt an, wie viel Lithium, ohne Betrachtung von Co-Kationen, maximal gebunden werden kann (Battistel et al. 2020). Da die Ausgangsmaterialien für die Elektroden sehr ähnlich zu den Adsorbern sind, liegen auch die maximalen Beladungskoeffizienten in einem ähnlichen Bereich von 30 – 40 mg/g bei Manganoxiden unter Laborbedingungen (Battistel et al. 2020; Liu, Zhao, and Ghahreman 2019; Calvo 2019). Bei einer Ionenpumpe wird in einem ersten Schritt eine Reaktionszelle mit dem lithiumführenden Fluid befüllt. Durch Anlegen einer Spannung wird das Lithium an die Arbeitselektrode gebunden und Chlorid an die Gegenelektrode, die z.B. aus Nickel oder Silber besteht (Battistel et al. 2020; Liu, Zhao, and Ghahreman 2019; Romero, Llano, and Calvo 2021; Zhao et al. 2019). Im nächsten Schritt wird die Sole aus der Kammer gespült und durch eine Rückgewinnungslösung ersetzt, in die in einem dritten Schritt das gebundene Lithium durch eine Spannungsumkehr freigegeben wird (Battistel et al. 2020; Zhao et al. 2019).

Die Elektrodialyse werden elektrochemische und membranbasierte Methoden verbunden. Zusätzlich zu den zwei Elektroden werden anionen- und kationenselektive Membrane verwendet (Stringfellow and Dobson 2021; Liu, Zhao, and Ghahreman 2019; X. Li et al. 2019; Mroczek et al. 2015). Diese trennen bei Anliegen der Spannung selektiv Lithiumkationen und Chloridanionen aus dem Ausgangsfluid ab. Je nach Konfiguration können so Anionen oder Kationen angereichert werden (Mroczek et al. 2015). In einem zweiten Schritt kann über eine Spannungsumkehr der zu gewinnende Rohstoff wieder an eine Trägerlösung abgegeben werden (Liu, Zhao, and Ghahreman 2019; Battistel et al. 2020; X. Li et al. 2019).

Da die elektrochemischen Abscheidemethoden auf den gleichen Mechanismen basieren wie sie bereits heute in Lithium-Ionen-Batterien zum Einsatz kommen, sind die Prozesse sehr gut verstanden (Liu, Zhao, and Ghahreman 2019). Durch das Anlegen der Spannung ist weiterhin, je nach Methode, die Adsorptionszeit deutlich niedriger (< 20 Minuten) als bei der Verwendung von Adsorbern allein (Battistel et al. 2020). Das Anlegen der Spannung ermöglicht zudem, dass keine Chemikalien für die Rücklösung oder Regeneration benötigt werden. Je nach Methode schwankt die benötigte Energie zwischen 1 und 60 Wh/mol ohne Berücksichtigung der Pumpenenergie (Battistel et al. 2020; Liu, Zhao, and Ghahreman 2019). Skaliert man diese Laborergebnisse rechnerisch auf ein Kraftwerk mit 4 MW installierter elektrischer Leistung, einer Fließrate von 80 L/s und 200 mg/L Lithium auf die Produktion einer Stunde um, erhält man 8300 mol Lithium pro Stunde (ca. 0.31 t/h LCE). Für eine vollständige Extraktion ergäbe das einen Energieverbrauch von 8,3 kWh – 498 kWh. Das Entspricht einem Anteil von 0,2 % - 12,4 % der in derselben Zeit produzierten Energie.

Im Vergleich der elektrochemischen Methoden geht man von einer höheren Effektivität der Ionenpumpe aus (Battistel et al. 2020). Die Elektrodialyse wurde zwar bereits erfolgreich im kleinen Maßstab mit geothermalen Fluiden getestet (Mroczek et al. 2015), jedoch stellt das Skalieren auf einen Anlagenmaßstab eine Herausforderung mit Blick auf die Haltbarkeit der selektiven Membranen dar (X. Li et al. 2019). Ein Problem, das alle Methoden teilen, ist die maximale Beladungskapazität der Arbeitselektroden. Für einen effektiven Einsatz in einem Geothermiekraftwerk ist daher das Design einer Reaktionszelle von größter Relevanz (Calvo 2019; Battistel et al. 2020; Zhao et al. 2019). Die korrosive Natur der Thermalwässer ist weiterhin eine Gefahr für die Integrität der Elektroden und kann die Leistungsfähigkeit beeinträchtigen (Calvo 2019). Zudem wurden bei elektrochemischen Wasseraufbereitungsprozessen pH-Wertschwankungen festgestellt, welche unkontrollierte Mineralausfällungen auslösen können (Obata et al. 2020; Dykstra et al. 2017; Arulrajan et al. 2021; Mroczek et al. 2015). Besonders das Anreichern von zweiwertigen Kationen wie $Ca^{2+}$, $Mg^{2+}$, $Fe^{2+}$ und Karbonaten im Allgemeinen kann hier die Leistungsfähigkeit und Langlebigkeit beeinträchtigen (Arulrajan et al. 2021). Beim Einsatz von Membranen in der Elektrodialyse besteht zusätzlich die Gefahr, dass diese durch Ablagerungen verstopft werden (X. Li et al. 2019; P. Zhang et al. 2018).

Trotz vieler offenen Punkte, wird den elektrochemischen Methoden ein sehr großes Potential für eine kommerzielle Nutzung im Anlagenmaßstab zugesprochen, vor allem aufgrund der kurzen Beladungszeit sowie des geringen Chemikalienbedarfs (Liu, Zhao, and Ghahreman 2019; Battistel et al. 2020; Calvo 2019; Zhao et al. 2019). Basierend auf den ersten erfolgreichen Laborversuchen, wird der Technologie-Reifegrad zwischen 3 und 4 geschätzt.

### 3.4 Membrantechnologien

Wie die Elektrodialyse basieren die Membrantechnologien auf dem Einsatz von selektiven Membranen, welche nur Lithium diffundieren lassen. Die selektive Trennung kann dabei über die Ionengröße, Oberflächenladung, oder chemische und physikalische Eigenschaften erfolgen (Stringfellow and Dobson 2021). Dabei gibt es eine große Variation an Membrantechnologien, wobei die Membran selbst die Trennung hervorruft, oder der Träger für Lösungsmittel oder Sorbentia sein kann (X. Li et al. 2019).

Nanofiltration, als Beispiel einer Membrantechnologie, ermöglicht die selektive Abscheidung multivalenter von monovalenten Ionen (X. Li et al. 2019; Y. Li et al. 2019; S. Y. Sun et al. 2015). Besonders bei magnesiumreichen Fluiden bietet sich eine Behandlung mittels

Nanofiltration an, da Mg bei vielen Methoden mit Lithium aufgrund ähnlicher Ionenradien konkurriert. Die Abscheidung erfolgt in einem druckbetrieben Prozess über die Porengröße sowie die Ladung der Membran (X. Li et al. 2019; Y. Li et al. 2019; S. Y. Sun et al. 2015). Nanofiltrationsanlagen sind bereits im industriellen Maßstab verfügbar und werden auch zur Wasseraufbereitung eingesetzt. Die Technik wurde auch schon mit Blick auf die Lithiumextraktion an komplexen Fluiden getestet (X. Li et al. 2019; Y. Li et al. 2019; S. Y. Sun et al. 2015). Alleine dienen sie jedoch nicht zur direkten Lithiumextraktion, da andere einwertige Kationen wie Natrium oder Kalium ebenfalls die Membran durchdringen. Daher muss der Einsatz der Nanofiltration zur Lithiumabscheidung immer in Kombination mit einer Vor- oder Nachbehandlung zur Abtrennung der Störionen erfolgen (X. Li et al. 2019; Y. Li et al. 2019; S. Y. Sun et al. 2015; Somrani, Hamzaoui, and Pontie 2013).

Ein weiterer Ansatz ist die Kombination von Membrandestillation und Kristallisatoren. Die Membrandestillation ist ein temperaturgetriebenes Verfahren, welches über eine wasserabweisende Membran kontaktlos eine Aufkonzentration bis hin zur Kristallisationsgrenze ermöglicht (X. Li et al. 2019; C A Quist-Jensen, Macedonio, and Drioli 2016; Cejna Anna Quist-Jensen et al. 2016). Der temperatur-betriebene Prozess ermöglicht dabei für geothermale Systeme die Perspektive einer netto-energieneutralen Wasserbehandlung. Des Weiteren verbraucht die Membrantechnologie kein Wasser. Es lässt sich im Gegenteil Frischwasser als Co-Produkt gewinnen (X. Li et al. 2019; Macdonio 2015). Ähnlich wie die Nanofiltration, ermöglicht die Membrandestillation nur eine Anreicherung der Mineralphasen bis hin zur Sättigungsgrenze. Für die finale Extraktion, muss ein separater Prozess oder Kristallisator verwendet werden(X. Li et al. 2019; C A Quist-Jensen, Macedonio, and Drioli 2016; Cejna Anna Quist-Jensen et al. 2016).

Bei der Methode der „Supported Liquid Membran" werden Membrane mit Lösungsmitteln beladen, welche für die Flüssig-Flüssig-Extraktion verwendet werden (X. Li et al. 2019; Sharma et al. 2016). Dadurch verbinden sich die Vorteile der Flüssig-Flüssig Behandlung mit einem deutlich verminderten Lösungsmittel- und Platzbedarf. Unter Laborbedingungen wurde dieser Ansatz für die Lithiumextraktion mit niedrigsalinaren Lösungen getestet und erreichte in Kreislaufprozess innerhalb 120 Minuten bei pH-Werten von 9,5 bzw. 12.5 Extraktionsraten von über 90 % (Sharma et al. 2016; Ma, Chen, and Hossain 2000).

Analog können auch Membran mit Ionensieben bestückt werden, was eine anlagentechnische Anwendung erleichtert (X. Li et al. 2019). Dadurch wird die große Oberfläche und Selektivität der Adsorber mit den Vorteilen der Membran kombiniert, welche den Adsorber stationär hält und einen niedrigen Energieverbrauch gewährleistet (X. Li et al. 2019). Die mit den Adsorbern bestückten Membranen konnten Beladungen von 30 mg/g erreichen, ähnlich wie die Manganadsorber alleine. Der größte Teil der Beladung, findet bei diesem Ansatz innerhalb von 60 Minuten Kontaktzeit statt und zeigt damit eine vergleichsweise schnelle Kinetik (D. Sun et al. 2016; X. Li et al. 2019).

Die Membranprozesse, können in der Rohstoffgewinnung aus Thermalwässern unterschiedliche Anwendungen finden. Nanofiltration und Membrandestillation sind effektive Methoden zur Aufkonzentration der Wässer und sind über ihren druck-, bzw. temperaturgetriebenen Prozess sehr gut in einen geothermischen Kreislauf integrierbar. Durch die Aufkonzentration können auch niedrigsalinare Wässer auf einen Lithiumgehalt gebracht werden, der eine

Rohstoffextraktion ermöglicht. Beide Methoden wurden auch schon zur Wasseraufbereitung im Anlagenmaßstab getestet. Die Membrantechnologien zur direkten Lithiumabscheidung, welche Adsorber oder Lösungsmittel mit Membranen kombinieren, sind im Vergleich dazu noch in einem Entwicklungsstadium. Eine selektive Abscheidung konnte bereits im Labormaßstab erfolgreich durchgeführt werden, jedoch bedarf es noch der Skalierung auf einen Anlagenmaßstab (Sharma et al. 2016; Ma, Chen, and Hossain 2000; D. Sun et al. 2016; Y. Li et al. 2019; X. Li et al. 2019). Der Technologie-Reifegrad liegt dabei zwischen 3 und 4. Durch die Kombinationen konnte die Leistungsfähigkeit der einzelnen Extraktionsmethoden in manchen Bereichen verbessert werden, Herausforderungen wie beispielsweise die Rücklösung des Lithiums mit Säuren, oder der Abbau von Adsorbern oder Lösungsmitteln bleiben aber bestehen. Hinzu kommen neue, membranspezifische Herausforderungen wie das Verstopfen der Membranen durch mineralische Ausfällungen in den Poren. Hier hebt sich jedoch das kontaktlose Verfahren der Membrandestillation ab, welches ein geringeres Potential für Verstopfungen zeigt. Weitere Herausforderungen sind die Komplexität der Herstellung der Membranen sowie die damit verbundenen hohen Kosten (X. Li et al. 2019).

## 4 Aktuelle Extraktionsprojekte

Weltweit wird in verschiedenen Regionen, wo Salare oder Thermalwässer mit hohen Lithiumkonzentrationen zu finden sind, an Techniken zur direkten Lithiumextraktion gearbeitet. In der Vergangenheit stand dabei die schnellere und erhoffte preisgünstigere Gewinnung von Lithium aus Solen im Vordergrund. Beispiele sind die Arbeiten von Livent in Catamarca (Argentinien) und von verschiedenen Firmen in Qinghai (China) (Grant 2020). Pilotarbeiten von Livent (früher FMC) gehen auf die 1990er Jahre zurück (Grant 2020). Eine kommerzielle Produktion mit direkter Lithiumgewinnung mittels Sorptionstechnik (Sorbentia wurde nicht spezifiziert) gibt es seit den 2000er Jahren in Catamarca (Argentinien). Versuche, diese Technologie auf Thermalwässer zu übertragen gab es zuerst in der Salton Sea Region (Imperial Valley, USA). Dabei spielte z.B. das Unternehmen Simbol Materials eine wichtige Rolle (Stringfellow and Dobson 2021). Durch die Volatilität des Marktes, standen die Firmen unter starkem finanziellen Druck und mussten häufig ihre Projekte einstellen. Durch die jetzt weiter gestiegene und v.a. projizierte, nachhaltig zunehmende Nachfrage und die aktuell hohen Lithiummarktpreise (s. Abb. 3) sind neue Akteure, wie zum Beispiel Energy Source Minerals hinzugekommen, die nach eigenen Aussagen z.T. bereits 2024 mit dem Beginn der Produktion jenseits des Pilotstadiums rechnen (Energy_Source_Minerals 2021).

### 4.1 Forschungsprojekte – Fokus Europa:

In Europa wird die direkte Lithiumextraktion aus Thermalwässern von wissenschaftlicher Seite (EuGeli, UnLimted) und durch verschiedene Unternehmen (Vulcan Energy, EnBW, Eramet) vorangetrieben. Eine Extraktion aus Salaren bzw. eine vorhergehende Konzentration der Thermalwässer steht nicht im Fokus. Es wird ein breites Spektrum verschiedener Technologien auf ihre Anwendbarkeit und Wirtschaftlichkeit untersucht. Im Folgenden gehen wir auf die einzelnen Ansätze in Europa und insbesondere in Deutschland ein.

Aktuell gibt es in Mitteleuropa drei laufende Forschungsprojekte zur Lithiumextraktion aus Thermalwässern:

### 4.1.1 UnLimited

UnLimited ist ein aktuell laufendes Forschungsprojekt eines Forschungsverbundes der EnBW mit dem Karlsruher Institut für Technologie (KIT), der Universität Göttingen, Bestec und Hydrosion und wird vom Bundesministerium für Wirtschaft und Energie mit 2.7 Millionen € gefördert. Nach eigenen Angaben könnten aus dem im Geothermiekraftwerk Bruchsal (Oberrheingraben) theoretisch extrahierten Lithium bis zu 20 000 Autobatterien pro Jahr hergestellt werden. In einer Pilotanlage soll Lithium mittels selektiver Sorbentia gewonnen werden. Die Konsortiumspartner untersuchen verschiedene Materialien und haben u.a. Manganoxide als vielversprechende anorganische Sorbentia ausgewählt (Unlimited 2021).

### 4.1.2 EuGeLi

Das EuGeLi Projekt ist ein Verbund aus neun Partner (ERAMET IDeas, BASF, BRGM, Chimie ParisTech, EIfER, Electricité de Strasbourg, IFPEN, VITO, Vrije Universiteit Brussel) und begann 2019 mit einer Laufzeit bis Ende 2021 (EuGeLi 2021). Das Budget wurde von EIT Raw Materials bereitgestellt. In einem Pilotprojekt am Geothermiekraftwerk Rittershoffen (Oberrheingraben) konnte Lithium aus dem Thermalwasser direkt gewonnen werden (Eramet 2021). Der verwendende Prozess bezieht sich auf eine nicht näher genannte Sorbens, das bei Betriebsdruck und –temperatur im Reinjektionsbereich zur selektiven Extraktion von Lithium aus dem Fluidstrom verwendet wird. Die Rückgewinnung des Lithiums aus dem Sorptionsmittel erfolgt als ein niedrig-salinares Eluat und orientiert sich an Extraktionsprozessen in Argentinien. Im Rahmen von EuGeLi wurde die Verwendung des Sorbens erfolgreich auf die Betriebsbedingungen des Reinjektionsstroms der Geothermiekraftwerke im Oberrheingraben angepasst und getestet (Rettenmaier et al. 2021).

### 4.1.3 BGR Projekt Li+fluids

Die Bundesanstalt für Geowissenschaften und Rohstoffe (BGR) leitet ein Forschungskonsortium zum Thema der Lithiumgewinnung aus Thermalwasser, das von BMWi gefördert wird (Stechern 2021). Das Projekt soll in seiner dreijährigen Laufzeit das Potential der Lithiumgewinnung evaluieren und die bisher bekannten Extraktionsverfahren bewerten. Zusätzlich sollen Lithiumfreisetzungsraten aus dem Tiefengestein zusammen mit einer Nutzwertanalyse bei der Co-Produktion im geothermischen Betrieb bestimmt werden.

## 4.2 Kommerzielle Projekte:

Im Oberrheingraben sind mehrere Unternehmen bezüglich der Entwicklung und Umsetzung von Techniken zur direkten Lithiumextraktion aus Thermalwässern aktiv. Sichtbar sind die Aktivitäten von Vulcan Energy. Des Weiteren sind aber auch die EnBW, Energie de Strasbourg (Es), GeoLith, Adionics und Eramet in Studien zu diesem Kontext involviert. International erscheinen auch die Projekte am Salton Sea (USA) weit vorangeschritten.

### 4.2.1 Vulcan Energy Resources

Die australisch-deutsche Firma Vulcan Energy setzt in ihrer Strategie auf ein Sorptionsmedium (Aluminiumhydroxid) (Wedin and Harrison 2021). Ein Vorbehandlungsschritt zur Entfernung von konkurrierenden Ionen geht der Extraktion mittels Sorbentia voraus (Wedin and Harrison 2021). Für die Extraktion wurden bereits mehrere kommerziell verfügbare Sorbentia im Rahmen von Laborstudien untersucht. Ein Demonstrator wurde bereits fertiggestellt, der 2022 mit der direkten Lithiumextraktion beginnen soll (Wedin and Harrison 2021). Ziel ist die Bereitstellung eines sehr reinen Lithiumchloridkonzentrats für die weitere Verarbeitung. Für die zukünftige Umsetzung der technischen Entwicklung in Richtung kommerzieller Lithiumextraktion hält Vulcan Energy mehrere Explorationslizenzen im Oberrheingraben. Eine kommerzielle Lithiumproduktion ist für 2024 vorgesehen (Vulcan_Energy_Resources 2021).

### 4.2.2 Geolith (Sustainable Lithium Solutions)

Geolith ist ein französisches Unternehmen, das sich auf die Extraktion von Lithium aus Solen und Thermalwässern fokussiert und verspricht, ab 50 mg Lithium/L Thermalwasser eine ökonomische Produktion (Geolith 2021). 2019 wurde die Forschung an der Lithiumgewinnung aus dem Thermalwasser vom französischen Institut ADEME (Agence de l'environment et de la Maitrise de l'Energie) sowie Mines Paris Tech unterstützt. Seit Beginn 2021 steht eine mobile Testanlage zur Verfügung, die für Tests an verschiedenen Geothermieanlagen gemietet werden kann. Auf Mikrofaser fixierte Sorbentia wirken selektiv zur Lithiumextraktion (Geolith 2021).

### 4.2.3 Adionics

Adionics ist ein 2012 gegründetes Unternehmen, das sich auf Entsalzungstechniken spezialisiert hat. Adionics hat eine Technik basierend auf den Prinzipien von „Thermal Swing Salt Absorption" (Variante der Flüssig-Flüssig Extraktion) entwickelt und patentiert, die eine selektive Extraktion von Salzen aus Solen ermöglicht. Möglich wird damit die direkte Gewinnung von LiCl aus Salaren und geothermalen Fluiden (Adionics 2021).

### 4.2.4 Internationale Projekte: Beispiele am Salton Sea (USA)

In der Salton Sea Region im Süden Kaliforniens ist eine größere Anzahl von Geothermiekraftwerken in Betrieb und damit bestehen gute Voraussetzungen für die Extraktion von Lithium (und anderen Elementen) aus dem Thermalwasser. Bereits 1976 wurde das dortige Thermalwasser als eine wichtige Ressource bzw. Reserve für Lithium betrachtet (Bertold and Baker 1976). In den 1970er Jahren wurden Laborstudien zur Lithiumextraktion durchgeführt, wobei die Ausfällung in Form von Lithiumaluminatkomplexen nach vorherigen Verdunstungskonzentration präferiert wurde (Bertold and Baker 1976). Des Weiteren wurden auch Versuche zur Lithiumgewinnung mittels Lithiumkarbonaten durchgeführt (Palmer, Howard, and Lande 1975). Tatsächlich umgesetzte Projekte zur Mineralextraktion haben sich in der Vergangenheit auf andere Elemente konzentriert, z.B. Kalium- und Calcium sowie Mangan und Eisen (1965-1990, verschieden Firmen) oder Zink (CalEnergy, um das Jahr 2000)(Stringfellow and Dobson 2021). SimbolMaterials stellte in den 2010er Jahren eine kommerziell skalierte Lithiumproduktion aus geothermischen Fluiden fertig, wurde aber durch Insolvenz an einem weiteren Vorantreiben des Projekts gehindert. Aktuelle Projekte werden von Energy Source Minerals am John L. Featherstone Geothermiekraftwerk mittels Sorbentia und von Controlled Thermal Resources am geplanten Hells Kitchen Geothermiekraftwerk vorangetrieben, jeweils mit prognostizierten Kapazitäten von 20.000 t LiOH (31.000 t LCE) pro Jahr. Des Weiteren gewann Berkshire Hathaway Renewables eine 6 Millionen US$ Ausschreibung und plant damit einen Demonstrator für LiOH-Gewinnung bis Frühjahr 2022 fertiggestellt zu haben. Nach erfolgreicher Umsetzung der Demonstrationssysteme ist eine kommerzielle Produktion bis 2024 anvisiert. Im technischen Bereich arbeitet Lilac Solutions an einem Ionenaustauschmaterial für hoch-selektive Lithiumextraktion mit Controlled Thermal Resources zusammen. Seit 2020 wird die Lithiumgewinnung im kalifornischen Imperial Valley/Salton Sea von staatlicher Seite durch eine "Lithium Valley" Kommission unterstützt (Lithium_Valley 2021).

# 5  Diskussion

## 5.1  Nutzung bestehender Kraftwerksanlagen

Betrachtet man die Entwicklungen des Lithiummarktes, so deuten unterschiedliche Szenarien auf ein sehr starkes Wachstum des Lithiumbedarfs hin. Dabei überschreiten die aktuellen Zahlen der Lithiumproduktion bereits die positivsten prognostizierten Szenarien von 2017. Mit dem Übertreffen dieser Produktionsszenarien ist ein Kriterium erfüllt, das einer Studie der DERA (Schmidt 2017) zufolge zu einem Lithiumangebotsdefizit am Markt im Jahre 2025 führen kann. Beim Eintreten dieser Situation ist von einem starken Anstieg des Preises für Lithium auszugehen. In Anbetracht eines wahrscheinlich drohenden Defizits und steigender Preise können auch alternative Lithiumreserven wie geothermische Wässer (wirtschaftlich) relevant werden. Unter Anbetracht des geplanten Ausbaus der Batteriezellfertigung in Deutschland wäre eine heimische Lithiumgewinnung unter anderem vorteilhaft für Lieferketten, Transportwege, Umwelt und könnte Deutschland eine gewisse Unabhängigkeit vom Weltmarkt und somit strategische Vorteile verschaffen.

Die hohen Lithiumgehalte in Thermalwässern stellen in Kombination mit den großen, Wasservolumina in den Reservoiren, theoretische eine vielversprechende Lagerstätte dar. Um eine mögliche Rohstoffproduktionskapazität abzuschätzen, ist die reine Betrachtung des Produkts aus Lithiumkonzentration im Wassers und der Fließrate eines Geothermiekraftwerks unzureichend. Dies führt in seinen vereinfachten Annahmen zu einer deutlichen Überschätzung des Potentials. Um zukünftig extrahierbare Mengen Lithium aus einer bestehenden Bohrung realistischer abschätzen zu können, müssen folgende reduzierende Faktoren mit einbezogen werden:

- Betriebsstunden des Geothermiekraftwerks: Selbst die zuverlässigsten Kraftwerke produzieren maximal 95% der Zeit eines Jahres. Für manche Standorte mit Scaling- und Korrosionsproblemen infolge einer komplexen Fluidzusammensetzung müssen zum Teil deutlich geringere Auslastungen angesetzt werden.
- Nutzbares Produktionsvolumen: Gerade in Anlagen mit großem Gewinnungspotential werden Volumenströme von 70 – 80 L/s produziert. Die Extraktionsmethoden benötigen Zeit und große Mengen Extraktionsmittel in Form von Sorbentia, Lösungsmitteln oder Wasser. Der Platz- und Infrastrukturbedarf steigt mit dem Volumenstrom und muss mit den Bedingungen einer geothermischen Anlage in Einklang gebracht werden. Aus diesem Grund, kann eine Extraktion eventuell nur auf einem Teilstrom realisiert werden.
- Lithiumkonzentration über die Zeit: Die geothermische Nutzung tiefer Fluide basiert in Deutschland auf Dublettensystemen (oder ähnlichen Konfigurationen). Aus Tracertests lässt sich über den Wiedererhalt an vielen Standorten ableiten, dass nicht unerhebliche Teile des Fluids mehrfach gefördert werden (Egert et al. 2020; Sanjuan, Scheiber, et al. 2016). Im Extraktionsbetrieb würde dies zu kontinuierlich abnehmenden Lithiumkonzentrationen im Fluid führen.
- Rohstoffextraktionsrate: Für keine Methode ist eine vollständige Extraktion zu erwarten. Bei Laborvalidierungen wurden vielfach bis zu 90% Lithium zurückgewonnen. Bei der Skalierung und Integration in den Kraftwerksbetrieb, sind hier zum Teil hohe Abschläge aus folgenden Gründen zu erwarten: komplexe Fluidzusammensetzung mit konkurrierenden Ionen, zu kurze Verweilzeiten im Extraktionssystem, hohe Volumenströme, hohe Temperaturen, u.a.m.

Tabelle 2: Lithiumkonzentration im Fluid, produzierter Volumenstrom und theoretisch extrahierbare Lithiummenge für Geothermiekraftwerke im Oberrheingraben und im norddeutschen Becken (Lithiumkonzentrationen aus (Sanjuan, Millot, et al. 2016; Regenspurg, Milsch, and Schaper 2015; Schallenberg 1999; Naumann 2000) Volumenströme aus (Dilger et al. 2021; Egert et al. 2020; Maurer et al. 2020))

| Geothermiekraftwerk | C(Li) Fluid [mg/L] | Q [L/s] | Extrahierbare Li-Menge bei Effizienz: 50% / 90% LCE [t/a] | Marktwert Worst-Case: 50 % Extraktion bei 9.000 US$/t LCE (GERMANLITHIUM 2021) | Marktwert Best-Case: 90 % Extraktion bei 50.000 US$/t LCE (GERMANLITHIUM 2021) |
|---|---|---|---|---|---|
| Bruchsal | 163 | 28 | 345 / 620 | $ 3.105.000 | $ 31.000.000 |
| Insheim | 168 | 80 | 1015 / 1826 | $ 9.135.000 | $ 91.300.000 |
| Landau | 181 | 70 | 957 / 1722 | $ 8.613.000 | $ 88.600.000 |
| Soultz sous Forêts | 173 | 30 | 392 / 705 | $ 3.528.000 | $ 35.250.000 |
| Rittershoffen | 190 | 70 | 1004 / 1807 | $ 9.036.000 | $ 90.350.000 |
| Groß Schönebeck | 215 | 15 | 243 / 438 | $ 2.187.000 | $ 21.900.000 |
| Neustadt-Glewe | 10 | 35 | 26 / 48 | $ 234.000 | $ 2.400.000 |
| Waren | 2.7 | 17 | 3 / 6 | $ 27.000 | $ 300.000 |
| Neubrandenburg | 1.8 | 28 | 4 / 7 | $ 36.000 | $ 350.000 |
| SUMME | | 469 | 3989 / 7229 | $ 35.901.000 | $ 361.450.000 |

Für die Berechnung der aus momentan installierten Anlagen in Oberrheingraben und dem Norddeutschen Becken extrahierbaren Mengen LCE (Abbildung 5) haben wir sehr optimistische Annahmen zugrunde gelegt. Die theoretisch extrahierbaren Massen sind in Abhängigkeit der Fließrate unter Produktionsbedingungen und der Lithiumkonzentration des Fluids ermittelt. Zusätzlich haben wir zwei unterschiedlich Extraktionseffizienzen von 50% und 90% als best- und worst-case Szenario betrachtet (Tabelle 2). Weitere Annahmen sind die Verfügbarkeit des Thermalwasserstroms 90%, ein verfahrenstechnisch nutzbarer Fluidanteil von 100% und eine über den Produktionszeitraum konstante Lithiumkonzentration. Für Insheim lässt sich abhängig von der Extraktionseffizienz so eine maximal extrahierbare Menge LCE von etwa 1000 bis 1800 t/a ermitteln. Ganz ähnliche Mengen ergeben sich für Rittershoffen (1000 bis 1800 t/a) und Landau (950 bis 1720 t/a). Aufgrund der niedrigeren Volumenströme in Soultz-sous-Forêts und Bruchsal können hier potentiell nur etwa 390 bis 700 t/a beziehungsweise 350 bis 620 t/a extrahiert werden. Am Standort Groß Schönebeck mit der höchsten Lithiumkonzentration (215 mg/L) führt ein geplanter Volumenstrom von nur 15 L/s so zu einer Lithiummenge von lediglich 240 bis 440 t/a. Die drei anderen Standorte in Norddeutschland (Neustadt-Glewe, Waren und Neubrandenburg) weisen alle Lithiumkonzentration zwischen 1 und 10 mg/L auf. Dies führt zu extrahierbaren Mengen von <50 t/a LCE.

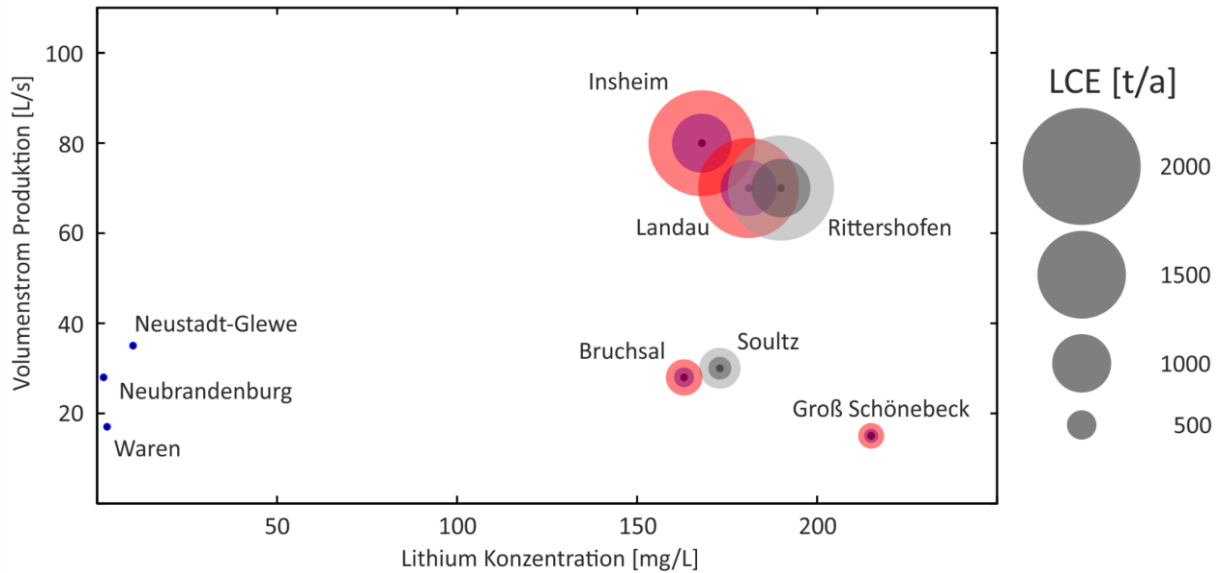

*Abbildung 5: Theoretisch extrahierbare Menge LCE in Tonnen pro Jahr für Extraktionseffizienzen von 50% (blau) und 90% (rot). Weitere Annahmen der Potentialermittlung sind eine konstante Lithiumkonzentration über den Produktionszeitraum, nutzbarer Fluidanteil 100% und eine Verfügbarkeit der geothermischen Produktion von 90%.*

Der Vergleich der unterschiedlichen Technologien zeigt, dass eine selektive Extraktion von Lithium aus geothermalen Wässern technologisch möglich ist und im Labor validiert wurde. Ein Transfer auf einen Prototyp oder Anlagenmaßstab ist dabei jedoch keinesfalls trivial. Die prognostizierte Extraktionseffizienz von 90 % sind in Laborversuchen nur mit langen Verweilzeiten erreicht worden. Diese sind in einem Geothermiekraftwerk aufgrund der großen Volumenströme jedoch nicht einzuhalten, vorausgesetzt, dass man aus dem gesamten Volumenstrom extrahieren möchte. Für eine Implementierung wird man daher geringere Extraktionsraten zu Gunsten schnellerer Kinetik in Kauf nehmen müssen, nur einen Teilstrom im Kraftwerk nutzen können oder einen deutlich höheren Materialeinsatz benötigen. Selbst die für das Erreichen von 50 %, der bei unterschiedlichen Methoden benötigten 60-minütigen Verweilzeit, ist mit Blick auf Fließraten von 80 L/s eine große Herausforderung. Dies gilt insbesondere, da die normale Verweilzeit des Thermalwassers im Geothermiekraftwerk bei ca. 5 Minuten liegt.

Weiterhin wird eine Extraktion zusätzlich durch das Korrosions- und Scalingpotential der Wässer erschwert. Diese sind bereits Hindernisse für die Energiegewinnung (Scheiber, Seibt, et al. 2019; Scheiber, Hettkamp, et al. 2019; Eggeling et al. 2018; Nitschke et al. 2014). Diese Effekte dürften sich infolge von Prozessen zur Fluidkonzentration, Abkühlung während der Reaktionszeit oder pH-Wert-Änderungen zur Verbesserung der Extraktionseffektivität, noch verstärken. Wie sich diese jedoch im Detail auf eine Rohstoffextraktion auf Produktionsskala auswirken, kann nur im Rahmen eines in-situ Langzeitanlagentest bestimmt werden. Punktuelle Prototypentests geben darüber nur eine bedingte Aussage. Von z.B. Grant, 2020, wurde hervorgehoben, dass ein einfacher Transfer der Erkenntnisse von einer Anlage auf andere Standorte aufgrund unterschiedlicher Thermalwasserzusammensetzung nicht uneingeschränkt möglich ist. Extraktionseffizienzen bleiben deshalb aufgrund fehlender großskaliger Demonstratoren spekulativ. Beim heutigen Stand der Technik erscheinen Extraktionsraten von 50 % im laufenden Betrieb allerdings realistischer als vielfach diskutierte, optimistischere Szenarien. Bei dem 50 % Effizienzszenario ergäbe sich für alle Standorte, auch die in Frankreich und die mit niedrigen Lithiumgehalten ein Gesamtproduktionsvolumen von ca. 4000 t LCE pro Jahr (bei 90% Verfügbarkeit des Geothermiekraftwerks, 100% Volumenstromnutzung). Das entspräche 3 - 11 % des Bedarfs für die in Deutschland geplanten Batteriezellfertigungen. Je nach eintretender Preisprognose (9000, 25.600, 50.000 US$/t LCE) ergibt sich damit ein Marktvolumen von 35, 100 oder 196 Mio. US$. Bei einer globalen Produktion von 420.000 t Lithiumkarbonat-Äquivalenten (LCE) in 2020 (Abbildung 1) könnte der Anteil der möglichen extrahierbaren LCE Menge aus den aufgeführten Geothermiekraftwerken etwa 1 % zum Weltmarkt beitragen und wäre damit in der gleichen Größenordnung wie aktuell die gesamte europäische Produktion (Kavanagh et al. 2018). Dabei handelt es sich um ein global messbares Produktionsvolumen und damit für Deutschland und Europa um einen relevanten Zubau. Jedoch befindet sich die deutsche Industrie im Wandel von einem Markt der fertige Batterien einkauft, hin zu einer eigenen Batteriezellfertigung. Dabei steht der geplante Ausbau der Zellfertigung einer nicht existenten Lithiumproduktion gegenüber, was eine Abhängigkeit vom globalen Lithiummarkt darstellen wird. Eine schnelle Implementierung der Lithiumgewinnung aus Thermalwässern erscheint allerdings aufgrund der Vorlaufzeiten für Geothermiekraftwerke, von der Exploration über das Genehmigungsverfahren bis hin zum Bau und dem Betrieb (durchschnittlich etwa 7-9 Jahre), schwierig. Für eine heimische Lithiumproduktion ist dieser Ausbau jedoch essenziell, um weitere lithiumreichen Reservoire zu erschließen und damit der größte Hebel für eine Lithiumproduktion aus Thermalwässern in Deutschland.

## 5.2   Reservoir- und Ressourcenbewertung

Ein wesentlicher Aspekt für eine Investition in die Extraktion werthaltiger Elemente aus Thermalwasser ist das erschließbare Wasservolumen in Kombination mit dem Lithiumgehalt. Im Gegensatz zum klassischen Bergbau ist ein schematisiertes Vorgehen für die Bewertung salzhaltiger Thermalwässer als Rohstoffvorkommen bislang nicht vorhanden. Einschätzungen für die Ressourcengröße bzgl. Thermalsolen wurden z.B. von Vulcan Energy vorgenommen. Für das Lizenzgebiet Ortenau wurde eine Ressourcengröße von 13.2 Millionen Tonnen LCE abgeschätzt (Vulcan Energy, 2019), wobei die Güte der Einschätzung der Kategorie mit der höchsten Unsicherheit („inferred ressource") für eine identifizierte Ressource entspricht. Eine Bewertung erfolgte über eine Klassifikation nach dem Australian Joint Ore Reserves Committee Code (JORC) aus dem Jahr 2012. Die aktuelle JORC Klassifikation bezieht sich primär auf die Bewertung von Mineralressourcen und Erzvorkommen und ist damit möglicherweise nicht optimal für die Einschätzung von Thermalwasserressourcen geeignet. Eine fundierte Abschätzung der Ressourcengröße eines Thermalwasserreservoirs wird zusätzlich erschwert, da viele reservoirspezifische Parameter zur Bewertung bekannt sein müssen und z.T. noch in laufenden Forschungsprojekten evaluiert werden müssen:

- Volumen des Fluidreservoirs
- Porosität

- Natürliche und künstlich veränderte Permeabilität im Aquifer
- Mögliche Aquitarde und verheilte Klüfte, die die Permeabilität reduzieren oder verhindern
- Klüfte und Störungszonen, die eine Verbindung zu darunterliegenden und darüber liegenden Aquiferen darstellen können und einen Austausch zwischen den Schichten ermöglichen
- Details der natürlichen Tiefenzirkulation
- Verbindung des geothermischen Kreislaufs zur Tiefenwasserzirkulation (welcher Anteil des Thermalwasserreservoirs steht damit in Verbindung? Zu welchem Grade steht die Produktions- und Injektionsbohrung in einem Kurzschlussverhältnis?)
- Herkunft des Lithiums und eine mögliche Wiederanreicherung im Reservoir

Für einen ökonomischen Betrieb einer Extraktionsanlage sind neben der Ressourcengröße auch das zeitlichen Verhalten der Fluidchemie von wesentlicher Bedeutung. Besteht ein hydraulischer Kurzschluss zwischen Produktions- und Injektionsbohrung, in dem nach wenigen Wochen reinijizierte Wasservolumina an der Produktionsseite gefördert werden, dann wird die Lithiumkonzentration bei laufender Extraktion in kurzer Zeit (i.e., Wochen oder Monate) abnehmen und die Produktion unwirtschaftlich und machen. Eine gute Anbindung an den Thermalwasseraquifer ist somit essentiell, mit einem nicht dominanten Anteil des reinfizierten Wassers an der Produktionsseite. Für die Standorte Soultz-Sous-Forêts und Rittershofen wurde bereits über Tracer-Tests nachgewiesen (Egert et al. 2020; Sanjuan, Scheiber, et al. 2016), dass das reinjizierte Wasser zu signifikanten Teilen wieder an der Produktionsbohrung gefördert wird.

In diesem Zusammenhang ist der Aspekt einer mögliche Lithium-„Nachlieferung" durch Lösung in Folge von Wasser-Gesteins-Wechselwirkung im Reservoir ungeklärt. Hierzu gibt es bislang aufgrund fehlender aktiver Extraktion keine in-situ Erfahrung. Eine Einschätzung ist nur aufgrund von Laborexperimenten und damit verbundenen Simulationen in sehr vereinfachter Weise möglich, da sie die Verhältnisse im Aquifer nicht vollständig widerspiegeln können. In der Studie von Drüppel et al. (2020) wurde z.B. das Laugungsverhalten von Graniten bei Exposition in salinaren Fluiden bei 70 °C und bei 200 °C mit 2 molarer NaCl Lösung untersucht. Bei 70 °C wird Lithium u.a. durch die Auflösung von Biotit und Muskovit ins Fluid eingebracht. Bei höheren Temperaturen um 200 °C spielt v.a. die Feldspatauflösung eine dominante Rolle. Bei Experimenten bei 200 °C mit Granit und Monzonit hatte die Laugungslösung nach dem 36 Tage dauernden Experiment eine Lithiumkonzentration von 1-2 mg/L. Dieser Wert zeigt die Größenordnung der zu erwartenden Lithiumkonzentration auf, wenn ein hypothetisch lithiumfreies Fluid nach Injektion für einen ähnlichen Zeitraum durch den Thermalwasseraquifer fließt und danach wieder gefördert wird. Für eine genauere Betrachtung müssen die Fließwege und Kontaktzeiten mit dem Reservoirgestein betrachtet werden.

Allgemein bedarf es auch detaillierter Untersuchungen und Modelle für das Verhalten des behandelten Wassers im geothermischen Reservoirs, welches von Lithium und eventuell bei der Extraktion hindernden Ionen befreit ist. Da die Wässer initial im Gleichgewicht mit dem Chemismus des Reservoirgesteins sind, wird dieses bei einer selektiven Extraktion gestört. Bei einer Reinjektion wird sich dieses Gleichgewicht durch chemische Gesteins-Wasser-Wechselwirkungen einstellen, was zu Mineralausfällungen oder Auflösungen im Thermalwasseraquifer führen kann und sich entsprechend auf die Nachhaltigkeit der Produktion auswirken kann.

# 6 Gesamtfazit

Die weltweite Lithiumnachfrage wird auch weiterhin stark ansteigen. Der Hauptreiber ist der steigende Bedarf an Lithium-Ionen-Hochleistungsspeichern, insbesondere zum Ausbau der E-Mobilität. Dabei wird sich der deutsche Lithiummarkt angesichts der Planungen zum Bau von Batterieproduktionsstätten an mindestens neun heimischen Standorten stark verändern. Der bisherige Import von bereits gefertigten Batterien, wird durch einen Rohstoffimport ersetzt werden müssen. Allein die vorgesehene Batteriefertigung würde mit geplanten 37.000 - 149.000 t LCE knapp die Hälfte der gesamten Weltproduktion aus 2020 erreichen. Bislang ist die globale Lithiumproduktion schlecht diversifiziert. Mehr als 80 % der gesamten Produktion stammen aus zwei Ländern, Australien und Chile. Dort ist, unabhängig von der Lagerstätte, die Gewinnung mit Umweltbeeinträchtigungen verbunden, welche durch die langen Transportwege bis zu den Batteriezellfertigungen noch vergrößert werden. Aus diesen Faktoren lässt sich eine starke geostrategische Bedeutung für eine deutsche Binnenquelle für Lithium ableiten.

Lithium ist auch in Deutschland in großen Mengen in tiefen geothermalen Fluiden gelöst. Die Konzentrationen an relevanten Standorten im Oberrheingraben liegen zwischen 160 und 190 mg/L. Im Norddeutschen Becken erreichen die Konzentrationen in der Bohrung Groß Schönebeck bis zu 215 mg/L. In Fluiden des Rotliegend wurden vereinzelt sogar knapp 400 mg/L nachgewiesen. Der Zugang zu diesen Vorkommen kann bislang nur durch Geothermiebohrungen erfolgen. Die installierte Kapazität der Geothermie könnte momentan nur in geringem Umfang beitragen den wachsenden deutschen Lithiumbedarf zu decken. In Deutschland sind an lithiumrelevanten Standorten (ORG und NDB) lediglich sieben produzierenden Tiefbohrungen aktiv. Kumuliert wird hier ein Volumenstrom von etwa 270 L/s gefördert. Daraus kann in einem sehr optimistischen Prognoseszenario (Effizienz 90%, Verfügbarkeit 90 %, konstante Lithiumkonzentration im geothermalen Fluid und einen nutzbaren Fluidanteil von 100 %) potentiell etwa 4600 t/a LCE produziert werden. Dies entspricht etwa 3 - 12 % des jährlichen Bedarfs aus der geplanten deutschen Batteriezellenproduktion. Es handelt es sich hierbei um eine rein theoretische Potentialabschätzung. Für die tatsächliche Machbarkeit ist immer noch die Extraktionstechnologie selbst der Angelpunkt. Im Labormaßstab wurden eine Reihe Methoden, mit mehrheitlich guter Effizienz von zum Teil >90 %, erfolgreich validiert. Jedoch fehl nach aktuell verfügbarer Datenlage bislang die erfolgreiche Skalierung zum industriellen Prozess. Verfügbare Dokumentationen belegen einen Technology Readiness Level von etwa TRL 4-5. Aus der Kommunikation einzelner Player im Oberrheingraben, in denen von Vorort-Demonstratoren berichtet wird, lässt sich ein TRL von 6 ableiten. Von einem kommerziellen Einsatz ist die Branche demnach noch ein gutes Stück entfernt.

Sollte der Durchbruch zum industriellen Prozess gelingen, müssen für die Umsetzung noch folgende standortspezifische techno-ökonomische Schlüsselfragen geklärt sein:

- Wie groß ist das Reservoir und wie nachhaltig kann es bewirtschaftet werden?
- Wie verhält sich die Lithiumkonzentration über die Zeit während der Produktion?

- Welche Extraktionsmethode macht aufgrund der Fluidchemie am meisten Sinn?
- Sind die gegenläufigen Schlüsselparameter Extraktionseffizienz und übertägige Fluidverweilzeit am Standort optimierbar? Ist dieses Design in den Anlagenprozess integrierbar?
- Wie groß sind die Volumenströmen, die sich aus den verfahrenstechnischen Behandlungen ergeben. Sind diese zu bewältigen? Welcher Materialeinsatz zu welchem Preis ist notwendig?
- Welche Infrastrukturen müssen aufgebaut werden?
- Müssen Betriebsparameter der Anlage (Temperatur, Druck) für die Extraktion angepasst werden?
- Wie hoch ist der Energiebedarf?
- Wie entwickelt sich das Scalingpotential in allen Anlagenkomponenten durch die Kopplung der Lithiumextraktion?
- Welchen Einfluss hat die Extraktion (evtl. mit Vorbehandlung) auf den Chemismus des Fluids? Wie wirkt sich dies auf das Reservoir und die Anlagenkomponenten aus?
- Welche Stoffe werden dem Fluid zugegeben (Inhibitoren, Extraktionsmedien, etc.) und entfernt (Rohstoff, Co-Präzipitat, Scaling)? Ist dies mit der rechtlichen Lage vereinbar?
- Welche Stoffe und Mengen fallen zur Entsorgung an?

Aufgrund der Komplexität der Thematik, der herausragenden Bedeutung der zukünftigen Rohstoff-Versorgung, sowie des zunehmenden Interesses der Gesellschaft, sollte dieser Prozess weiterhin wissenschaftlich begleitet werden. Eine verstärkte Kooperation zwischen Industrie und Forschung wäre für einen effizienten technologischen Fortschritt sehr konstruktiv.

Diese Zusammenstellung basiert auf aktuellen wissenschaftlichen Publikationen. Interne Arbeiten und Patentanträge sind den Autoren weitgehend unbekannt und sollten in einer nächsten Studie evaluiert werden. Eine regelmäßige Bewertung der Situation erscheint auch durch den zu erwartende dynamischen Forschungsfortschritt notwendig. Die Autoren und die verantwortlichen Lehrstühle am KIT wollen hiermit ein "White Paper" initiieren, in dem laufend weitere wissenschaftliche Analysen eingearbeitet werden können.

# 7 Danksagungen

# 8 Literaturverzeichnis